\documentclass[11pt]{article}
\usepackage{epsf}
\usepackage{epsfig}
\usepackage[T1]{fontenc}
\usepackage{amsmath}
\usepackage{amssymb}
\usepackage{graphicx}
\usepackage{color}

\newcommand{\Ft}{\overline{\phi}_{t_0,t_0+\tau}}
\newcommand{\Fz}{\overline{\phi}_{0,\tau}}

\newcommand{\noi}{\noindent}
\newcommand{\Wt}{\overline{\Omega}_{t_0,t_0+\tau}}
\newcommand{\Wz}{\overline{\Omega}_{0,\tau}}

\newcommand{\be}{\begin{equation}}
\newcommand{\ee}{\end{equation}}
\newcommand{\bea}{\begin{eqnarray}}
\newcommand{\eea}{\end{eqnarray}}

\newcommand{\p}{{\bf p}}
\newcommand{\q}{{\bf q}}

\newcommand{\ie}{\it i.e.}
\newcommand{\qdot}{\dot{\bf{q}}_i}
\newcommand{\pdot}{\dot{\bf{p}}_i}

\newcommand{\za}{\alpha}

\newcommand{\zg}{\gamma}
\newcommand{\zG}{\Gamma}
\newcommand{\zd}{\delta}

\newcommand{\ze}{\varepsilon}

\newcommand{\zL}{\Lambda}

\newcommand{\zS}{\Sigma}
\newcommand{\zt}{\tau}

\newcommand{\zR}{I\hskip-3.6pt R}

\newcommand{\zW}{\Omega}
\newcommand{\WFR}{$\Omega$-FR }
\newcommand{\WFRs}{$\Omega$-FR's }

\begin{document}

\title{The Steady State Fluctuation Relation for the Dissipation Function}

\author{Debra J. Searles%
\thanks{Nanoscale Science and Technology Centre and School of Biomolecular and Physical Sciences, Griffith
University, Brisbane, Qld 4111, Australia (D.Bernhardt@griffith.edu.au)%
}, Lamberto Rondoni%
\thanks{Dipartimento di Matematica and CNISM, Politenico di Torino, Corso
Duca degli Abruzzi 24, 10129 Torino, Italy (lamberto.rondoni@polito.it) %
}  and Denis J. Evans%
\thanks{Research School of Chemistry, Australian National University, Canberra,
ACT 0200, Australia (evans@rsc.anu.edu.au)%
}}

\date{\today}

\maketitle

\newpage{} \noindent\textbf{Abstract}\\
 \\
We give a proof of transient fluctuation relations for the entropy production (dissipation function) in nonequilibrium systems, which is valid for most time reversible dynamics. We then consider the conditions under which a transient fluctuation relation yields a steady state fluctuation relation for driven nonequilibrium systems whose transients relax, producing a unique nonequilibrium steady state.  Although the necessary and sufficient conditions for the production of a unique nonequilibrium steady state are unknown, if such a steady state exists, the generation of the steady state fluctuation relation from the transient relation is shown to be very general. It is essentially a consequence of time reversibility and of a form of decay of correlations in the dissipation, which is needed also for, e.g., the existence of transport coefficients.  
Because of this generality the resulting steady state fluctuation relation has the same degree of robustness as do equilibrium thermodynamic equalities.  The steady state fluctuation relation for the dissipation stands in contrast with the one for the phase space compression factor, whose convergence is problematic, for systems close to equilibrium. We examine some model dynamics that have been considered previously, and show how they are described in the context of this work.
 \\

 \newpage{}


\vskip 25pt 


\section{INTRODUCTION}

Steady state fluctuation relations provide information about the probability
distribution of time averages of certain phase variables for systems
in nonequilibrium steady states: they express the ratio of the probabilities
of observing positive and negative values of time averages of such
variables. These relationships have received much attention over the
past decade (see e.g. \cite{review,AustJChem,GGbook,Bustamante}).
In fact, they are among the few exact results regarding nonequilibrium
systems; close to equilibrium they lead to the Green-Kubo
relations for linear transport coefficients \cite{ESR}; they provide 
information about fluctuations in nanoscale systems and they show how 
irreversibility can emerge from time reversible dynamics.

The first such fluctuation relation (FR) for a nonequilibrium particle system
was given in Ref.\cite{ECM2}, in 1993. It concerned fluctuations
of time averages of the energy dissipation or equivalently
the phase space contraction rate, for an isoenergetic shearing interacting particle
system. In the isoenergetic dissipative system of  Ref.\cite{ECM2}, the 
instantaneous energy dissipation
equals the instantaneous phase space contraction rate. The
form of the relation was the following:
\be
\frac{P_\zt(A)}{P_\zt(-A)} = e^{\zt A}
\label{firstFR}
\ee
where $A$ and $-A$ were the time averaged values of the dissipation
on steady state trajectory segments of duration $\zt$, and
$P_{\tau}{\scriptscriptstyle {\displaystyle (A)}}$ is the steady
state probability of observing the average value A (with some tolerance), 
in a time $\zt$. In analogy with
the periodic orbit expansion for dynamical systems \cite{POEinECM2},
the relation was derived heuristically using the {}``Lyapunov weights'' 
in the long $\zt$ limit, and was not expected to work at short $\zt$'s.

Paper \cite{ECM2} motivated a number of works, in which various
derivations of formulae formally similar to Eq.(\ref{firstFR}) were
given for different systems, although for non-isoenergetic systems 
the phase space contraction rate and the energy dissipation are no 
longer instantaneously equal. Beginning in 1994, a series of papers 
by Evans and Searles focussed on the fluctuation properties of the
``dissipation function'' $\zW$, cf.\
Refs.\cite{review,ES1,ES2,generalized,SE2000,HeatFlow,StephenPRE,stephennew},
while Gallavotti and Cohen, in
1995 $\mathit{et.}seq.$, considered the fluctuations of the time-averaged
phase space contraction rate, $-\zL$, under the assumption that the
dynamics is time reversible, transitive and Anosov or that the system
satisfies the Chaotic Hypothesis ($\mathit{i.e.}$ although the dynamics
is not Anosov, for the purposes of this FR, it behaves as if it was 
Anosov), cf.\ Refs.\cite{GC,GG-MPEJ,GGrevisited}.
Various generalizations and extensions of these relations have been
produced by different authors, see e.g.\  
Refs.\cite{generalized,stochasticES,stochasticL,othertheory,GG-GNS,openTRV,AES01}.

In the literature, FR's
which superficially resemble the form of Eq.(\ref{firstFR}), are
often collectively referred to as the Gallavotti-Cohen Fluctuation
Relations, or instances of the Gallavotti-Cohen symmetry, although
they may have subtly different meanings. Therefore, for the sake
of clarity, we refer to the relation for the phase space
contraction rate of Refs.\cite{ECM2,GC}, as the $\zL$-FR, while
we refer to the relation for the dissipation function
of Refs.\cite{review,ECM2,ES1,ES2}, as the $\zW$-FR. In the cases
in which $\zW=-\zL$, like those of \cite{ECM2}, the two 
relations have the same content. 

$\zW$-FR's have been widely used and tested numerically and experimentally,
for various transient \cite{ES1,ES2} as well as steady state
systems (see e.g. \cite{AustJChem,SE2000,stephennew,stochasticES,AES01,romans,DK,exptss, PRV}). 
Although the transient \WFRs have been proven more than a decade
ago, beginning with Ref.\cite{ES1,ES2}, there has been some
question over the derivation of the corresponding steady state relations
\cite{CG99}. Therefore, we present a derivation of the $\zW$-FR's which
is more thorough than the ones that have appeared thus far in the
literature \cite{review}. The purpose is to shed light on the mechanisms
which allow the $\zW$-FR's to hold, and which may be used to better understand 
properties of nonequilibrium systems. In particular, we are
interested in nonequilibrium steady states where the dissipation function 
takes the form  
\be 
\Omega = \zS = -JVF_e/(k_B T) ~,
\label{epr} 
\ee
where $J$ is the dissipative flux in the system of interest, $V$ is the volume of the system of interest, 
$k_{B}$ is Boltzmann's constant, $T$ is the equilibrium thermodynamic temperature of the thermal reservoir 
with which the system of interest is in contact, and $\zS$ is the generalized entropy production (or energy dissipation) induced by the dissipative field $F_e$ applied to the system of interest \cite{review,SE2000}. For the equality $\Omega = \zS$ to hold, we require that 
the unthermostatted ({\ie} adiabatic) field dependent dynamics preserves phase space volumes 
instantaneously, and that the initial distribution of phases could in principle be 
generated by a single (exceedingly long) zero-field phase space trajectory.
Systems that satisfy the condition of the adiabatic incompressibility of phase space
are called $AI\Gamma$ in Ref.\cite{DenisGarybook} (cf.\ Appendix A). 

When referring to steady state systems,
we only consider systems which relax to a {\it unique} steady state.
The necessary and sufficient
conditions for the system of interest to relax to a unique steady state are
unknown. Certainly, it is known that under certain circumstances nonequilibrium
steady states are not possible \cite{LiemBrownClarke}. 
If the systems of interest are near equilibrium,
where the thermodynamic notions of temperature and entropy can be
used, then as is shown in the theory of linear irreversible thermodynamics the time or ensemble average of $\zW$ is equal to leading order in the dissipative field, to the average
so-called spontaneous entropy production rate \cite{DGM}.
Outside this local equilibrium regime, the entropy production rate
cannot be defined. 
While the notions of a thermodynamic temperature, thermodynamic entropy
or an entropy production rate cannot be defined outside the local
thermodynamic equilibrium regime, in a nonequilibrium steady state
the ensemble or time averaged phase space contraction rate
is equal to the steady state time derivative
of the fine grained Gibbs entropy. The fact that the fine grained
Gibbs entropy is not constant in a steady state (unlike the pressure
or the energy), is the reason why it is so difficult to attach meaning
to standard thermodynamic quantities such as temperature, in steady states outside
the local equilibrium regime.

In the derivation of a relation like (\ref{firstFR}), one may follow
different approaches. The one which has led to the $\zL$-FR requires
the full knowledge of the physical measure of the dynamics, which
is then assumed to be of the Anosov type, hence to have a Sinai-Ruelle-Bowen 
(SRB) measure, $\mu_{SRB}$  \cite{GC,DRdiff}. This is the 
mathematical approach, which aims to identify the class of dynamical systems 
for which a relation like (\ref{firstFR}) can be rigorously obtained, 
regardless of the physical relevance of such dynamical systems or of the 
quantities appearing in (\ref{firstFR}). How far one can go along this line
and which other condition could replace the the Anosov condition are highly 
nontrivial and interesting mathematical questions, which do not have an answer, 
at present.

The explicit knowledge of the physical measure gives the maximum possible 
amount of information about the phase space distribution:  all statistical 
properties of the dynamics can be obtained from it. However, if one is only 
interested in why the FR holds, less detailed knowledge of the phase space 
distribution is likely to suffice, and special features of the dynamics, 
such as the Anosov property, need not play a role. This may 
explain why many {\em physical} systems have an observable which obeys a relation 
like (\ref{firstFR}). 
As a matter of fact, hardly any particle system of physical interest
is of the Anosov kind, therefore it is interesting to ask whether a 
physical system obeying a fluctuation relation does so because it 
shares certain properties with Anosov systems, or whether it obeys such 
relations for other reasons.
To investigate this question, which is part of the
physical approach, we follow Evans and Searles \cite{SE2000}, since they  
do not require the invariant phase space probability measure to be explicitly
determined. This will lead to transient fluctuation
relations for $\zS$ as well as for many other functions of phase, including $\zL$.
If $\phi$ is the observable of interest, we speak of $\phi$-FR. 

The purpose of the present paper is to understand why the $\zS$-FR holds for 
many physically interesting systems, and why this can be observed within 
physically relevant time scales. This is useful to further develop the
present theory.

\vskip 5pt \noi
{\bf Remark.} {\it This issue is not completely unrelated to, but is 
quite different from the identification of the class of dynamical 
systems which allow a rigorous  mathematical derivation of the $\zL$-FR. 
One may say that the two issues are as much related as Khinchin's approach 
to the ergodic hypothesis \cite{Khin} is related to ergodic theory. 
}

\vskip 5pt
In Section 2 we derive transient FR's for time reversal invariant 
dynamics. These relations are called ``transient'' because 
they describe the time evolving statistics of an ensemble of finite time 
averages of given phase functions. 
They hold under very general conditions, similarly to other transient 
relations, like the Jarzynski and the Crooks relations \cite{JC}. 
In particular, the transient relations hold even if a steady state with
positive and negative fluctuations of the observable under consideration
is not reached. If a fluctuating steady state is not reached, the
transient relations hold and express properties of evolving ensembles of
phase space trajectories, while the Anosov structure cannot characterize
the dynamics.

In Section 3, we show how the results of Section 2 can be used to derive 
asymptotic FR's for reversible systems, under the technical assumption 
that $\zW$ is bounded. In Section 4, 
the decay of the $\zW$-autocorrelation is shown to be sufficient for the 
steady state \WFR to hold. 
Furthermore, in isoenergetic systems which enjoy $AI\zG$
(e.g.\ systems whose adiabatic evolution is Hamiltonian),
the phase space contraction rate and the dissipation 
function are instantaneously equal. In these cases, the predictions of the $\zL$-FR
and of the \WFR can be compared directly, and 
the \WFR at high dissipation implies a different behavior from the one 
predicted by the $\zL$-FR under the axiom-C condition of Ref.\cite{axiomC},
but which agrees with the results of \cite{stephennew,BGG}. This is possible because 
the conjectures of \cite{axiomC,BGG} are not necessarily verified by particle systems.
The approach followed in Sections 2, 3 and 4 leads to a number of
$\phi$-FR's and to various other relations.
In Section 5 we discuss our results. Appendix A contains a
detailed description of the dissipation function for the common cases.
Appendix B discusses the problem of axiom-C systems and the decay of correlations, and explains that our approach is based on exact relations, which are then valid even 
when the conditions of Section 4 are not met.  
Our conclusions are that: 

\begin{itemize}
\item to satisfy the steady state $\zW$-FR, the time reversal invariance of the dynamics 
and the decay of the $\zW$-autocorrelations, with respect to the {\em initial} 
(non-singular) probability measure, suffice. This decay of correlations is not the one 
characterizing the steady states of axiom-C systems, which concerns all observables and 
is referred to the {\em invariant} (singular) measure. 
\item For $\zW$ to equal $\zS$, hence to obtain a steady state 
$\zS$-FR, $AI\zG$ must hold and the
initial distribution of phases must in principle be obtainable form a single 
field free ($F_e=0$) phase space trajectory. In such a case, we say that 
the initial distribution is ergodically consistent with the field free dynamics \cite{review}.
\item The $\zW$-FR's avoid the difficulties encountered by the
$\zL$-FR close to equilibrium. The $\zL$-FR, requires ever longer
convergence times, the closer the
steady state is to equilibrium \cite{ESR,SE2000,romans,DK,BGGZ}. The 
$\zW$-FR's hold in a range which does not shrink for 
decreasing fields, 
and are verified within the physically relevant ($\zW$-autocorrelation)
times, both for systems far and close to equilibrium. 
\item The direct derivation of the steady state  $\zS$-FR's avoids the difficulties
connected with obtaining a relation for the energy dissipation
via the $\zL$-FR, justifies the convergence rates, and leads to new testable 
relations. However, the  
class of dynamical systems which enjoy the required decay of correlations
remains to be identified.
\item It goes without saying that the steady state relations require that a steady 
state is reached (hence that the correlations decay with respect 
to the initial state), and that fluctuations of opposite sign in the dissipation 
function should be observable. If such a steady state is not attained or if the 
steady state is sufficiently far from equilibrium that negative values of the 
dissipation function cannot be observed, the steady 
state FR's cannot be applied, but the transient FR's (including their asymptotic 
forms) remain valid as properties of the evolving ensembles, because they only 
require time reversibility. 
\end{itemize}
In this sense, our relations are like the usual thermodynamic relations which are 
extraordinarily general, being independent of the nature of interparticle 
interactions and dynamics.  All that is required of the dynamics is that 
it should be time reversible and, for steady state FR's, that the autocorrelation of $\zW$ decays and that the steady state is unique\footnote{This is analogous to the equilibrium thermodynamic requirement of a unique allotrope.}. Furthermore our derivation explains why the $\zW$-FR is verified within times which can be expressed in terms of material properties.

\section{TRANSIENT FLUCTUATION RELATION}
Because we will be interested in thermodynamic systems which possibly relax to a 
nonequilibrium steady state, our dynamics will consist of the Hamiltonian dynamics 
of an $N$-particle system of interacting particles to which a dissipative field 
is applied. {The resultant equations of motion} may or may not be derivable from a Hamiltonian 
\cite{DenisGarybook}. 
In order for such a system to be capable of relaxing to a nonequilibrium steady state, 
the system must be allowed on average, to loose heat.  This can be 
accomplished by surrounding the system of interest with thermostatting walls which 
serve two purposes: to confine the system of interest and also to remove dissipative 
heat generated (on average) by the dissipative field applied to it. 
It has been shown that any loss of heat from a Hamiltonian system 
requires that the dynamics over the degrees of freedom in the system of interest is 
no longer volume preserving \cite{DRPhysToday}. These details are explained 
further in the Appendix. In what follows however, we will consider a more general 
setting that is time reversible and does not necessarily preserve the phase space volumes.

Given the phase space ${\cal M}$ of a particle system, consider a
probability measure $\mu$ on ${\cal M}$, with density $f$, ${\ie}$ 
$d \mu(\zG) = f(\zG) d \zG$ for every point $\zG \in {\cal M}$. 
The measure
$\mu$ does not need to be produced by any dynamics on ${\cal M}$,
although it could represent, for instance, the initial equilibrium
distribution of the system. Further, given a (sufficiently well behaved)
phase function $\phi:{\cal M}\rightarrow\zR$, the probability that
it takes values in the interval $(a,b)\subset\zR$, according to $\mu$,
is given by
\be
\int_{\phi|_{(a,b)}} d \mu(\zG)
= \int_{\phi|_{(a,b)}} f(\zG) d \zG
\label{P_F}
\ee
where 
\be
\phi|_{(a,b)} {:=} \{ \zG \in {\cal M} : \phi(\zG) \in (a,b) \} ~,
\ee
is the set of points of the phase space for
which $\phi$ takes values in $(a,b)$.\footnote{$\phi|_{(a,b)}$ is 
assumed to be $\mu$-measurable, like all sets in this paper.}

Let $S^{\tau}:{\cal M}\rightarrow{\cal M}$ be the time evolution
operator, which takes any point ${\Gamma}\in{\cal M}$
to its evolved image $S^{\zt}\zG$, under some dynamics
applied for the time $\zt$.%
\footnote{In particular, the dynamics could be determined by the equations of
motion of a particle system subjected to a dissipative field $F_{e}$,
and to a deterministic thermostat meant to remove the excess energy
pumped into the system by the field, so that a nonequilibrium steady
state can be reached \cite{DenisGarybook}. In this case,
$\zG=(\mathbf{q{\textstyle {\textstyle ,}\mathbf{p}}})$ denotes the
generalized coordinates and momenta of all the particles comprising
the system. 
} We refer to time reversal invariant dynamics, {\ie}
dynamics which obey 
\be 
iS^\zt \zG = S^{-\zt} i \zG \quad \mbox{
for all } \zG \in {\cal M} \quad \mbox{and all } \zt \in
\zR \ee
where $i:{\cal M}\rightarrow{\cal M}$, which obeys $ii=i^{2}=$~identity,
is an involution representing the time inversion operator (e.g.\
$i{{\Gamma}}\equiv i(\mathbf{q},\mathbf{p})\equiv(\mathbf{q},-\mathbf{p})$)
for the dynamics. Consider the phase functions defined by: 
\be
\Ft(\zG) {:=} \frac{1}{\tau} 
\int_{t_0}^{t_0+\tau} \phi(S^{s} \zG) d s ~, \quad 
\phi_{t_0,t_0+\zt}(\zG) {:=} 
\zt \Ft(\zG) ~, \quad
\mbox{for } ~~
t_0,\zt \in \zR ~; \quad \zG \in {\cal M}
\label{phitau}
\ee
Take $\zd>0$, $t_{0}=0$,
introduce the sets $A_{\zd}^{+}=(A-\zd,A+\zd)$, $A_{\zd}^{-}=(-A-\zd,-A+\zd)$
and, for an odd phase variable $\phi$ ({\ie} $\phi(i \Gamma)=-\phi(\Gamma)$), consider the ratio
\be 
\frac{
\mu(\Fz|_{A^+_\zd})}{ \mu(\Fz|_{A^-_\zd})} = \frac{
\int_{\Fz|_{A^+_\zd}} f (\zG) d \zG }{ \int_{\Fz|_{A^-_\zd}} f (\zG) d \zG } ~, \label{POoverPO} \ee
{\ie} the
probability according to $\mu$ that $\Fz\in A_{\zd}^{+}$, divided
by the probability according to the same $\mu$ that $\Fz\in A_{\zd}^{-}$.

To compute this quantity, observe that the points of ${\cal M}$
which fall in $\Fz|_{A^-_\zd}$ are those, and only those,
obtained by doing the time inversion of the evolution, for a time
$\zt$, of the points in $\Fz|_{A^+_\zd}$, {\ie} 
\be 
\Fz|_{A^-_\zd} = i S^\zt \Fz|_{A^+_\zd} ~. \label{timeinversion}
\ee
Indeed, take any $\zG\in \Fz|_{A^-_\zd}$, invert it and
evolve it backward for a time $\zt$; the resulting point $Y=S^{-\zt}i\zG$
is the initial condition of a trajectory segment of duration $\zt$
over which one obtains $\Fz\in A_{\zd}^{+}$, hence $Y\in \Fz|_{A^+_\zd}$.
Moreover, there are no points of $iS^{\zt}\Fz|_{A^+_\zd}$
which do not lie in $\Fz|_{A^-_\zd}$. This allows us to compute
the denominator of Eq.(\ref{POoverPO}) in terms of $\Fz|_{A^+_\zd}$,
through the coordinate transformation 
\be 
\zG = i S^\zt X ~, \qquad
\mbox{whose Jacobian is } \quad J = \left| \frac{d \zG}{d
X} \right| =\exp\left( \int_0^\zt \zL(S^s X) d s \right) = e^{\zL_{0,\zt}(X)}
~. \label{ccordtransf} 
\ee
Here, the quantity $\zL$ is the phase
space expansion rate (the opposite of the contraction rate) which,
for dynamics $\dot{\Gamma}=G(\zG)$ on ${\cal M}$, is defined by
\be 
\zL(\zG) = \nabla \cdot \dot{\Gamma} = \nabla \cdot
G (\zG) ~. 
\ee
This leads to: 
\be 
\int_{\Fz|_{A^-_\zd}}
f (\zG) d \zG = \int_{\Fz|_{A^+_\zd}} f (i S^\zt X) \left|
\frac{d \zG}{d X} \right| d X = \int_{\Fz|_{A^+_\zd}}
f (iS^\zt X) \, e^{\zL_{0,\zt}(X)} d X \label{int-A} 
\ee
A necessary condition for the derivation of the transient fluctuation
relations is that \[
\int_{\Fz|_{A^-_\zd}}f(\zG)d\zG\ne0\quad\mbox{for all }A\mbox{ for which }~~\int_{\Fz|_{A^+_\zd}}f(\zG)d\zG\ne0\]
This condition is satisfied by appropriate choices of $\mu$, for the given 
$S^{\zt}$. For example, $\mu$ could be selected to be the probability 
distribution generated by the field-free (equilibrium) dynamics of a 
particle system, $S_0^\zt$ say, for which $S^{\tau}$ is the nonequilibrium
dynamics; {\ie} $\mu$ could be generated by $S_{0}^{\zt}$, obtained by 
setting the dissipative field of the nonequilibrium dynamics $S^{\zt}$ 
to zero. In fact, this selection ensures that 
$f(iS^{\tau}\Gamma)\ne0$ whenever $f(\Gamma)\ne0$.\footnote{Here,
it is assumed that equilibrium dynamics corresponds to no net
dissipation, and that typical trajectories explore all the phase space.}
Then, assuming that $f$ is also time reversal invariant\footnote{The 
assumption that $f$ be time reversal invariant is not necessary for our 
main results. A different choice will simply result in a alternative 
definition of the dissipation function, \cite{review}. 
However, if $\mu$ is selected to be an equilibrium 
measure, then this condition is guaranteed, and relying on it 
makes the calculations more elegant.} ({\ie}
that $f(\Gamma)=f(i\Gamma)$ for every $\zG\in{\cal {M}}$) the ratio
of probabilities (\ref{POoverPO}) becomes: 
\bea
\frac{\mu(\Fz|_{A^+_\zd})}{\mu(\Fz|_{A^-_\zd})}=
\frac{ \int_{\Fz|_{A^+_\zd}} f (\zG) d \zG }{ 
\int_{\Fz|_{A^+_\zd}} f (i S^\zt X) \exp\left( \zL_{0,\zt}(X) \right)
d X } =\frac{ \int_{\Fz|_{A^+_\zd}} f (\zG) d \zG }{
\int_{\Fz|_{A^+_\zd}} f (S^\zt X) \exp\left( \zL_{0,\zt}(X)
\right) d X } \label{POoverPO2} 
\eea
The restriction on the choice of $\mu$ for the given $S^{\tau}$,
{\ie} that $f(iS^{\tau}\Gamma)\ne0$ whenever $f(\Gamma)\ne0$,
is referred to as \textit{ergodic consistency} of $f$ with $S^\zt$ 
\cite{review}. 
The main quantity used below can now be introduced.

\vskip 10pt 
\noindent{\bf Definition.}{\it The time averaged
{\em dissipation function} $\Wt : \mathcal{M} \to \zR$, for a 
time reversal invariant
phase space probability density $f$ is defined by \cite{review}:

\bea
\Wt(\zG) {:=} \frac{1}{\zt} \int_{t_0}^{t_0+\zt} 
\zW(S^s \zG) d s & {:=} &\frac{1}{\zt}
\left[ \ln \frac{f(S^{t_0}\zG)}{f(iS^{t_0+\zt} \zG)} - 
\int_{t_0}^{t_0+\zt} \zL(S^s \zG) d s \right]
\nonumber \\
&=&\frac{1}{\zt}
\left[ \ln \frac{f(S^{t_0}\zG)}{f(S^{t_0+\zt} \zG)} - 
\zL_{t_0,t_0+\zt}(\zG) \right]
\label{omegat}
\eea
provided $f$ is ergodically consistent with $S^t$. The instantaneous
dissipation function $\Omega(\zG)$ is obtained from differentiation
with respect to $\zt$ of $\Wz$, for sufficiently regular $f$. }

\vskip 10pt 
\noindent 
Here the condition of ergodic consistency plays
an important role, although it may not be so evident in general. Indeed,
given a particle system with certain conserved quantities under its
equilibrium dynamics $S_0^\zt$, the nonequilibrium dynamics $S^\zt$ may 
not conserve the same quantities, and a trajectory $S^{\zt}\zG$ may wander 
outside the support of $\mu$, making the corresponding $\zW$ unusable in
the following derivations. Thus, in physical applications $f$ must 
be chosen with care, although many phase space densities are acceptable
from a mathematical point of view. Using Eq.(\ref{omegat}),
Eq.(\ref{POoverPO2}) implies what has been referred to as the

\noindent \vskip 10pt \noi \textbf{Transient ${\phi}$-FR:} 

\be
\frac{ \mu(\Fz|_{A^+_\zd})}{ \mu(\Fz|_{A^-_\zd})} =
\frac{ \int_{\Fz|_{A^+_\zd}} f (\zG) d \zG }{
\int_{\Fz|_{A^+_\zd}} 
\exp\left[-\zW_{0,\zt}(X) \right] f(X) d X } 
:=
\left\langle \exp \left( - \zW_{0,\zt} \right) \right\rangle_{\Fz \in A^+_\zd}^{-1}
\label{PooverPO3}
\ee

\vskip 10pt \noi Here $\left\langle \exp\left(-\zW_{0,\zt}\right)\right\rangle _{\Fz\in A_{\zd}^{+}}$
is the ensemble average of $\exp[-\zW_{0,\zt}(X)]$, over the set of trajectories
which satisfy the constraint that $\Fz\in A_{\zd}^{+}$ \cite{review,generalized}.
In the case that $\Fz=\Wz$ the fluctuation relation assumes the particularly
elegant form, called

\vskip 10pt \noi \textbf{Transient $\zW$-FR:} 
\be 
\frac{\mu(\Wz|_{A^+_\zd})}{\mu(\Wz|_{A^-_\zd})} =\left\langle \exp
\left( - \zW_{0,\zt} \right) \right\rangle_{\Wz \in A^+_\zd}^{-1}
=e^{{[}A+\ze(\zd,A,\zt)]\zt} ~, \label{PWoverPW} 
\ee

\vskip 10pt \noi Here, $\ze$ is an error term of magnitude $|\ze|\le\zd$,
which depends on $\zd,A$ and $\zt$, and appears because $\Wz$ is
not necessarily constant and equal to $A$. Therefore, $\ze(\zd,A,\zt)\rightarrow0$
as $\zd\rightarrow0$, for all $A$ for which the ratio of Eq.(\ref{PWoverPW})
exists \cite{review,ES1,ES2}.

\vskip 5pt \noi {\bf Remark.} {\it Different choices of $f$ are possible, which lead to different $\zW$-FR's. In particular, equilibrium probability densities of many-particle systems which obey 
$AI\zG$ lead to an $\zW$-FR which concerns the entropy production, or the energy dissipation 
rate $\zS$ (cf.\ Appendix A). The uniform density in a compact phase space, $f(\zG)=1/|{\cal M}|$ say, yields $\zW=-\zL$. A density such that $f d \zG$ approximates as accurately as desired the (possibly singular) steady state measure is also allowed.}

\noindent \vskip 5pt

The above relations are called ``transient'' because, independently
of the length of $\zt$, they express properties of the measure $\mu$
and not of the possible steady state. The time $\zt$, indeed, enters
the $\phi$-FR only in the definition of the observable $\Fz$, and
not of the phase space probability density. Equations (\ref{PooverPO3},\ref{PWoverPW})
are exact and do not need any limit to be taken either on $\zt$ or
on $\delta$. As the calculation shows, the only properties that are
required for Eqs.(\ref{PooverPO3},\ref{PWoverPW}) to hold are that
the dynamics $S^{\zt}$ be time reversal invariant, to ensure that,
for odd $\phi$, $-A$ can be observed if $A$ can, and that $f(iS^{\tau}\Gamma)\ne0$
for any $\Gamma$ for which $f(\Gamma)\ne0$. The range of $A$'s
that can be observed depends both on $S^{\zt}$ and $\phi$. 

If the initial distribution can be generated by a single field free phase 
space trajectory, and $AI\zG$ holds, the dissipation function $\zW$ takes
the form given in (\ref{epr}) and at equilibrium (where $F_{e}=0$),
$\zW=0$, hence
\be 
\frac{ \mu(\Fz|_{A^+_\zd})}{\mu(\Fz|_{A^-_\zd})} = 
\left\langle e^0 
\right\rangle_{\Fz \in A^+_\zd}^{-1} = 1 
\label{normalization}
\ee
as it should be for the equilibrium measure $\mu$, and for any
odd variable $\phi$ (for instance, $\phi$ could be the instantaneous
current $J$, which does not need to vanish at equilibrium, although
its mean certainly does). For symmetric intervals $(-\zd,\zd)$, one
also has $A=0$ and can write
\be 
\frac{ \mu(\Fz|_{(-\zd,\zd)})}{ \mu(\Fz|_{(-\zd,\zd)})} = 1 ~, \quad \mbox{hence } ~~ 
\left\langle e^{- \zW_{0,\zt}} \right\rangle_{\Fz \in (-\zd,\zd)} = 1 ~.
\label{interN}
\ee
In the limit that $\zd$ tends to infinity, the second equality in
(\ref{interN}) tends to a full phase space average, and produces the
NonEquilibrium Partition Identity  (also referred to as the Kawasaki normalization)\cite{ES2,DenisGarybook}.
This identity can be used to test the accuracy of the numerical simulations,
and has even been used to calibrate experimental equipment \cite{exptss}. The highly
asymmetric  
convergence of numerical estimates to unity means that estimating the statistical uncertainty of experiments  is problematic without this relationship \cite{ES2}.
If the dissipation function is bounded (i.e.\ $|\zW|\le\zW^{*}$, for 
some $\zW^{*}>0$), one obtains 
\be 
e^{-\zt \zW^{*}} \le \frac{
\mu(\Fz|_{A^+_\zd})}{ \mu(\Fz|_{A^-_\zd})} \le e^{\zt
\zW^{*}} ~, 
\label{bddd}
\ee
for all odd $\phi$. Because different dissipation functions are allowed
and the above ratio holds for all of them, a sharp bound can be given {\em for 
all} $\phi$ and independently of the physically relevant dissipation function,
by taking the smallest possible $\zW^{*}$.   
This constitutes a new prediction, which may 
be interesting to test in nonequilibrium systems.

\section{ASYMPTOTIC FLUCTUATION RELATIONS}
\begin{figure}[htbp] 
   \centering
   \includegraphics[width=5in]{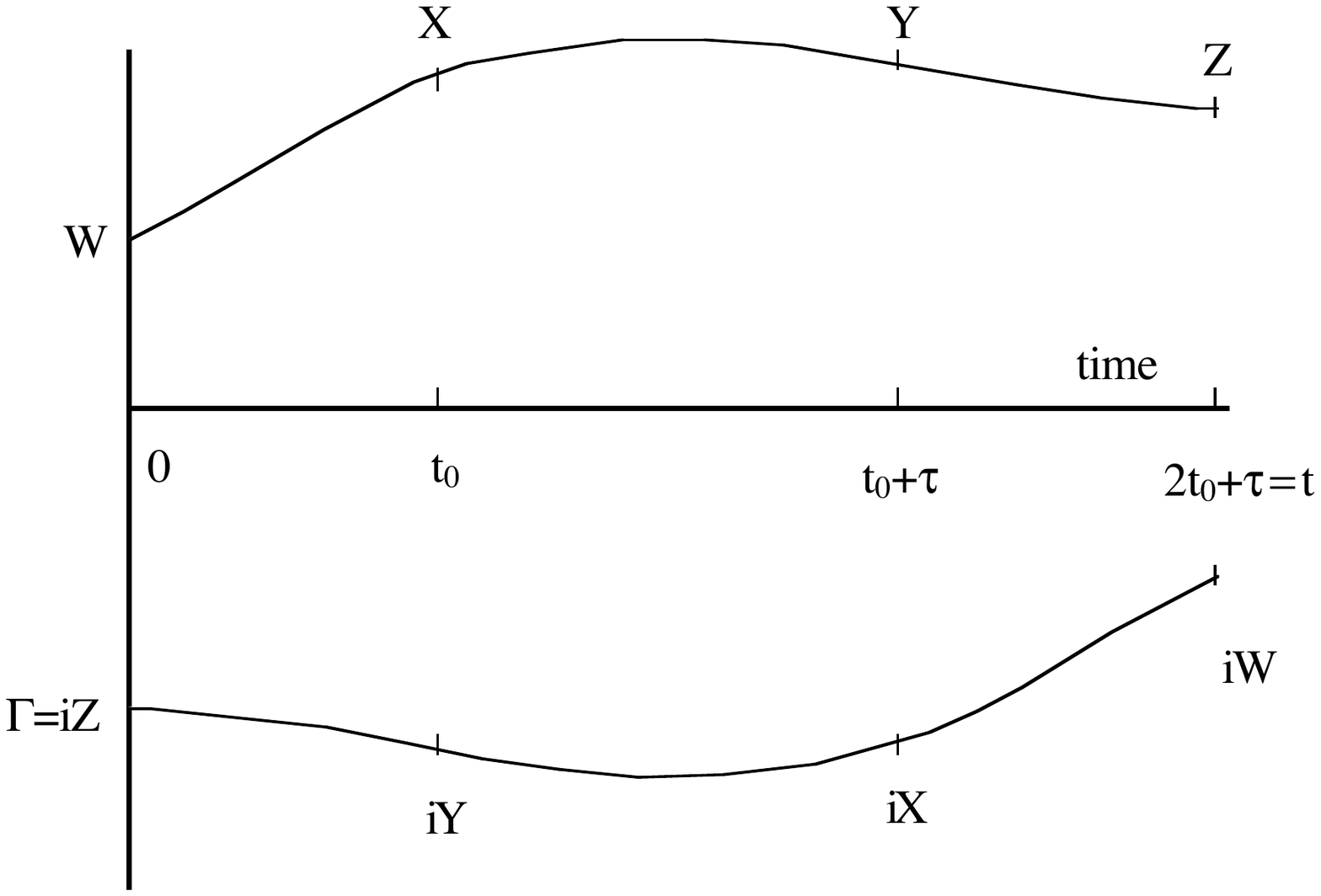} 
   \caption{Schematic diagram illustrating the time-reversible evolution of 
two points in phase space that are related by a time reversal mapping, $i$.
If the average of $\phi$ between $X$ and $Y$ is $A$, the average between
$iY$ and $iX$ is $-A$.}
   \label{XWYGZ}
\end{figure}

In order to derive steady state fluctuation relations, one may develop
further the transient $\phi$-FR's obtained in Section 2, considering
the time averaged phase variable $\Ft$ (cf.\ Eq.(\ref{phitau})),
where the time averaging begins at a time $t_{0}>0$, rather than
at time 0: 
\be 
\frac{\mu(\Ft|_{A^+_\zd}) }{\mu(\Ft|_{A^-_\zd}) } = 
\frac{ \int_{\Ft|_{A^+_\zd}} f (\zG)
d \zG }{ \int_{\Ft|_{A^-_\zd}} f (\zG) d \zG } ~. \label{PpoverPp1}
\ee
Here, $\Ft|_{E}=\{\zG\in{\cal M}:\overline{\phi}_{t_{0},t_{0}+\zt}(\zG)\in E\}$
is the set of all phase points which are initial conditions of trajectory
segments of duration $t_{0}+\zt$, over which $\Ft$ takes values
in $E$, and $E=A_{\zd}^{+}$ or $A_{\zd}^{-}$. The calculation of
the ratio (\ref{PpoverPp1}) requires that the points of ${\cal M}$
lying in $\Ft|_{A^-_\zd}$ be identified. To do that, consider
a trajectory of length $t=2t_{0}+\zt$, and take $W$ in $\Ft|_{A^+_\zd}$.
Then, let $X=S^{t_{0}}W$, $Y=S^{t_{0}+\zt}W$, $Z=S^{2t_{0}+\zt}W=S^{t}W$,
and $\zG=iZ=iS^{t}W$ (see Figure \ref{XWYGZ}). The trajectory segment between
$iY$ and $iX$ produces the opposite of the average produced between $X$ and
$Y$, in the time interval $(t_0,t_0+\zt)$, so that 
\be 
\Ft|_{A^-_\zd} = i S^{t} \Ft|_{A^+_\zd} \label{timeinversionp} 
\ee
and, as in the previous section, 
\bea 
\int_{\Ft|_{A^-_\zd}} f (\zG) d \zG &=& 
\int_{\Ft|_{A^+_\zd}} f (i S^t W) \left| \frac{d \zG}{d W} \right| d W \nonumber \\
 &=& \int_{\Ft|_{A^+_\zd}} f (S^t W) \exp\left( \zL_{0,t}(W)
\right) d W \label{int-phitau} 
\eea 
where $J=\left|\frac{d\zG}{dW}\right|=\exp[\zL_{0,t}(W)]$ and we use time reversal invariance of $f$.

Note that the integral defining $J$ is along the full trajectory
of duration $t$, although $\Ft$ concerns only one of its parts of
duration $\zt$. From the definition of $\overline{\Omega}_{0,t}$
(Eq.(\ref{omegat})), one obtains: 
\be 
\zL_{0,t}(W) = \ln \frac{f(W)}{f(S^t
W)} - \zW_{0,t}(W) \label{Lambdatagain} 
\ee
and 
\be 
\int_{\Ft|_{A^-_\zd}} f (\zG) d \zG = \int_{\Ft|_{A^+_\zd}} f (W)
e^{- \zW_{0,t}(W)} dW ~. \label{phiomegatau} 
\ee
The probability ratio of Eq.(\ref{PpoverPp1}) can then be written as:
\be
\frac{\mu(\Ft|_{A^+_\zd}) }{\mu(\Ft|_{A^-_\zd}) } = \left[
\frac{\int_{\Ft|_{A^+_\zd}} f (W) e^{- \zW_{0,t}(W)} d W}{
\int_{\Ft|_{A^+_\zd}} f (W) d W} \right]^{-1} 
:=
\left\langle \exp \left( - \zW_{0,t} \right) 
\right\rangle_{\overline{\phi}_{t_0,t_0+\zt} \in A^+_\zd}^{-1}
\label{PPoverPPSS}
\ee
which is the inverse
of the average of $\exp(-\zW_{0,t})$, conditioned to the constraint
$\overline{\phi}_{t_{0},t_{0}+\zt}\in A_{\zd}^{+}$; and the special 
case $\phi=\zW$, yields 
\be 
\frac{\mu(\Wt|_{A^+_\zd})}
{\mu(\Wt|_{A^-_\zd}) } = \left\langle \exp \left( - \zW_{0,t} \right)
\right\rangle_{\Wt \in A^+_\zd}^{-1} ~. 
\label{exactTFTOmega} 
\ee
The right hand side of this expression can be written as:
\be
\left\langle \exp \left( - \zW_{0,t} \right) \right\rangle_{\Wt \in
A^+_\zd}^{-1} = e^{\left[ A + \ze(\zd,t_0,A,\zt)\right] \zt}   
\left\langle e^{- \zW_{0,t_0} -  \zW_{t_0+\zt,2t_0+\zt}} 
\right\rangle_{\Wt \in A^+_\zd}^{-1}
\ee
Taking the logarithm, and dividing
by $\zt$ one then obtains: 
\be 
\frac{1}{\zt} \ln 
\frac{\mu(\Wt|_{A^+_\zd})}{\mu(\Wt|_{A^-_\zd})} = A + \ze(\zd,t_0,A,\zt)
- \frac{1}{\zt} \ln \left\langle e^{-\zW_{0,t_0} - \zW_{t_0+\zt,2t_0+\zt}}
\right\rangle_{\Wt \in A^+_\zd} \label{exactFT_t_0} 
\ee
where
the correction term $\ze$ depends on $\zd,t_{0},A$ and $\zt$ and,
like in Section 2, $|\ze|\le\zd$. This result is exact, and holds
for all $t_{0}$, $\zt$ and $\zd$, and for any $A$ for which both
the probabilities in the left hand side of (\ref{exactFT_t_0}) do
not vanish. It rests only on the time reversibility of $S^{t}$, and
on the ergodic consistency of $f$ with $S^t$. 

Equations (\ref{PPoverPPSS}) and (\ref{exactFT_t_0}) are
valid for all $\zt$, hence also in the limit $\zt\to\infty$, but
remain transient relations because no matter how large $\zt$ is;
they refer to the initial measure $\mu$, even if the dynamics makes
$\mu$ evolve into $\mu_{\zt}$. However,
the above derivation suggests that for systems that reach a steady
state, it may be possible to derive steady state FR's
from the transient FR's. The idea is that for sufficiently
large times $t_{0}$, Eqs.(\ref{PPoverPPSS},\ref{exactFT_t_0})
should concern trajectory segments of length $\zt$ distributed
according to the invariant measure, or very close to that. Furthermore,
taking $\zt$ long, and $\zd$ small, should make the last two 
terms in Eq.(\ref{exactFT_t_0}) negligible. 

To investigate this possibility, we firstly transform Eqs.(\ref{PPoverPPSS})
and (\ref{exactFT_t_0}) by transferring the time evolution up to $t_{0}$
from the sets of phase space points $\Ft|_{(a,b)}$, to the phase
space probability distributions. Given an initial phase space probability
density $f_0$, and the corresponding probability 
measure $\mu_0$, we denote their evolved counterparts after a time 
$t_{0}$ by $f_{t_{0}}$,
and $\mu_{t_{0}}$ respectively: 
\be
\mu_{t_0}(S^{t_0} E) = \mu_0(E)  \quad \mbox{or equivalently } ~~~
\int_{S^{t_0}E} f_{t_0}(X) d X = \int_E f_0(W) d W \label{consproft}  
\ee
for every $t_{0}\in\zR$
and $E\subset{\cal M}$, where $S^{t_{0}}E$
is the set of points reached by all points of $E$ after
an evolution of duration $t_{0}$. Equations (\ref{consproft}) are
consequences of the conservation of probability by time evolution.%
\footnote{They merely state that the initial probability of $E$ becomes
the probability of the evolved set $S^{t_{0}}E$ after a time $t_{0}$.%
} Physically, they are an instance of what is referred to as the {\em
Heisenberg-Schr\"{o}dinger equivalence}, while mathematically they are
the result of the coordinate transformations $X=S^{t_{0}}W$ and $W=S^{-t_{0}}X$.
The Jacobian of this transformation is given by 
\be 
\left| \frac{\partial
W}{\partial X} \right| = e^{\zL_{0,-t_0}(X)} = e^{\int_0^{-t_0}
\zL(S^s X) d s} = e^{-\int_{-t_0}^0 \zL(S^s X) d s} = e^{-\zL_{-t_0,0}(X)}
\ee
hence one can write 
\bea 
\mu_0(\Ft|_{(a,b)}) &=& 
\int_{\Ft|_{(a,b)}} f_0(W) d W = \nonumber \\
 &&\int_{S^{t_0} \Ft|_{(a,b)}} f_0 (S^{-t_0}X) e^{-\zL_{-t_0,0}(X)}
d X \label{variousp} 
\eea 
and 
\be 
\mu_0(\Ft|_{(a,b)}) = \mu_{t_0}(S^{t_0}
\Ft|_{(a,b)}) = \int_{S^{t_0} \Ft|_{(a,b)}} f_{t_0}(X)
d X ~. \label{muevolv} 
\ee
In Eqs.(\ref{variousp},\ref{muevolv}),
we could have integrated over any measurable set $E$, 
hence, equating (\ref{variousp}) and (\ref{muevolv})
finally gives:
\be 
f_{t_0} (X) = f_0(S^{-t_0} X) e^{-\zL_{-t_0,0}(X)}
~. \label{ft_0W} 
\ee
This is known as the Lagrangian form of the Kawasaki distribution function 
\cite{ES2}.\footnote{This can also be written, 
$f_{t_0} (S^{t_0}W) = f_0(W) e^{-\zL_{0,t_0}(W)}$.}
From Eq.(\ref{Lambdatagain}), which yields
\be
-\zL_{-t_0,0}(X)=\int_0^{-t_0} \zL(S^s X) d s = 
\ln \left(\frac{f(X)}{f(S^{-t_0} X)} \right) - \zW_{0,-t_0}(X) ~,
\label{intSsW}
\ee
Eq.(\ref{ft_0W})
can be rewritten as 
\be 
f_{t_0} (X) = f_0(X) e^{- \zW_{0,-t_0}(X)}
~, \label{ft_0WOmega} 
\ee
and gives the evolution of the phase space probability density $f_0$. 
This well known exact result for the time dependent phase space density 
$f_{t_{0}}(X)$, was first derived for thermostatted particle dynamics
where the dissipation function takes the form given in Eq.(\ref{epr}) by
Evans and Morriss in 1984 \cite{EM84}. Observe that 
\be 
S^{t_0} \Ft|_{(a,b)}
= \Fz|_{(a,b)} 
\ee
because $\phi$ is a phase variable, hence
does not explicitly depend on time, and because we assume the dynamics
to be autonomous. Indeed, a point $X\in S^{t_{0}}\Ft|_{(a,b)}$
is the evolution by a time $t_{0}$ of a point $\zG$ for which $\Ft$
takes the same value of $\Fz(X)$, i.e.\
\be
\Fz(S^{t_0} \zG) = \frac{1}{\zt} \int_0^\zt \phi(S^{s}(S^{t_0} \zG)) \, d s = 
\frac{1}{\zt} \int_{t_0}^{t_0+\zt} \phi(S^{s} \zG) \, d s = \Ft(\zG) ~.
\ee
Thus, Eq.(\ref{muevolv}) expresses the
time evolving probability of the fixed set $\Fz|_{(a,b)}$, 
\be
\mu_{t_0}(\Fz|_{(a,b)}) = \mu_{t_0}(S^{t_0} \Ft|_{(a,b)})
~. \label{probequiv} 
\ee
Provided convergence to the steady state occurs, this equation yields 
a more and more accurate approximation to the steady state measure of 
$\Fz|_{(a,b)}$, as $t_{0}$ grows, at fixed $\zt$.

Equation (\ref{PPoverPPSS}), which is expressed in terms of the initially 
chosen phase space probability distribution, and of evolving
phase space sets, can now be given in terms of the evolving PDF, since
\be
\frac{\mu_{t_0}(\overline{\phi}_{0,\zt}|_{A^+_\zd})}
{\mu_{t_0}(\overline{\phi}_{0,\zt}|_{A^-_\zd})} =
\frac{\mu_{t_0}(S^{t_0}\Ft|_{A^+_\zd})}{\mu_{t_0}(S^{t_0}\Ft|_{A^-_\zd})}
= \frac{\mu_0(\Ft|_{A^+_\zd})}{\mu_0(\Ft|_{A^-_\zd})} ~.
\ee
This produces what we call
\vskip 10pt \noi
{\bf $\phi$-FR:}
\be
\frac{\mu_{t_0}(\overline{\phi}_{0,\zt}|_{A^+_\zd})}
{\mu_{t_0}(\overline{\phi}_{0,\zt}|_{A^-_\zd})} = 
\langle \exp \left(
-\Omega_{0,t} \right) \rangle^{-1}_{\Ft \in A^+_\zd} \
\label{ESSFT}
\ee
Relation (\ref{ESSFT}) is exact and holds for
all $t_{0},\zt$ and possible pairs $A$ and $-A$ (whose range
must be determined case by case). Letting $\Ft=\Wt$ and rearranging,
Eq.(\ref{exactFT_t_0}) yields the

\vskip 10pt \noi 
\textbf{$\zW$-FR:} 
\be 
\frac{1}{\zt} \ln
\frac{\mu_{t_0}(\overline{\zW}_{0,\zt}|_{A^+_\zd})}
{\mu_{t_0}(\overline{\zW}_{0,\zt}|_{A^-_\zd})} = A +
\ze(\zd,t_0,A,\zt) - \frac{1}{\zt} \ln \left\langle e^{-\zW_{0,t_0}
- \zW_{t_0+\zt,2t_0+\zt}} \right\rangle_{\Wt \in A^+_\zd} ~,
\label{exactFTt_01} 
\ee

\vskip 10pt \noi where the conditional ensemble average in the last
term is computed with respect to the initial distribution $f$.

Let us consider some direct consequences of the 
above relations. 
For $A=0$, $\zd>0$, $t_{0} , \zt \in \zR$, and any odd $\phi$, the 
conditional average of Eq.(\ref{ESSFT}) yields
\be 
\langle
\exp \left( -\Omega_{0,t} \right) \rangle_{\Ft \in (-\zd,\zd)}
= \frac{\mu_{t_0}(\overline{\phi}_{0,\zt}|_{(-\zd,\zd)})}
{\mu_{t_0}(\overline{\phi}_{0,\zt}|_{(-\zd,\zd)})}
= 1 \label{inter} 
\ee
which, in the $\zd\to\infty$ limit, leads again to the NonEquilibrium 
Partition Identity (\ref{interN}), for the full ensemble average. 

Taking $A=0$ in Eq.(\ref{exactFTt_01}), one obtains
\be 
\langle e^{-\zW_{0,t_0}
- \zW_{t_0+\zt,2t_0+\zt}}\rangle_{\Wt \in (-\zd,\zd)}= e^{\ze(\zd,t_0,A=0,\zt)
\zt} \label{1stprop} 
\ee
which shows that this particular conditional average is uniformly bounded in 
$t_{0}$, and tends to 1 when $\zd \to 0$, at fixed $\zt$. Equation (\ref{inter}) 
with $\phi=\zW$ differs from (\ref{1stprop}) because of the integration times in 
the argument of the exponential function. For fixed $\zd$ and $\zt$, such that 
$\zd \zt \ll 1$, one can write 
\be 
\langle e^{-\zW_{0,t_0}
- \zW_{t_0+\zt,2t_0+\zt}} \rangle_{\Wt \in (-\zd,\zd)} \approx
1 ~, 
\ee
even for $t_0 \to \infty$, which is interesting for the validity of the steady 
state $\zW$-FR. Indeed, the last term in Eq.(\ref{exactFTt_01}) ought not to 
dominate over the others, if the steady state $\zW$-FR is to hold.  
Exchanging the roles of $A_{\zd}^{+}$ and $A_{\zd}^{-}$ in Eq.(\ref{exactFTt_01}),
one obtains 
\be 
\left\langle
e^{-\zW_{0,t_0} - \zW_{t_0+\zt,2t_0+\zt}} \right\rangle_{\Wt
\in A^-_\zd} = \frac{e^{\ze(\zd,t_0,A,\zt)+\ze(\zd,t_0,-A,\zt)}}
{ \left\langle e^{-\zW_{0,t_0} - \zW_{t_0+\zt,2t_0+\zt}}
\right\rangle_{\Wt \in A^+_\zd}} ~, 
\ee
for any $\zd>0$.

One last property that is worthwhile to consider is the following
asymptotic result, which is particularly close to a steady
state FR. As it happens in many situations of interest, let $|\zW|$ be
bounded by some $\zW^{*}>0$, so that: 
\be 
\left| \ln \left\langle
e^{-\zW_{0,t_0} - \zW_{t_0+\zt,2t_0+\zt}} \right\rangle_{\Wt
\in A^+_\zd} \right| \le {2 t_0 \zW^{*}} 
\label{boundedW} 
\ee
for all $\zt$. Then, taking $\zd<\zg$, which implies $|\ze|<\zg$, and 
\be 
\zt(t_0) \ge \frac{2 t_0 \zW^{*}}{\zg - \zd} = \zt^*(t_0) ~, 
\label{taucondition}
\ee
one obtains the following

\vskip 10pt \noi 
\textbf{$\zW_\infty$-FR:} 
\be
A - \zg \le \lim_{t \to \infty} \frac{1}{\zt(t)}
\ln \frac{\mu_{t}(\overline{\zW}_{0,\zt(t)}|_{A^+_\zd})}
{\mu_{t}(\overline{\zW}_{0,\zt(t)}|_{A^-_\zd})} \le A + \zg
\label{almostSS}
\ee

\noi
Recalling the connection
between $\zW$ and the physical dissipation, which holds under proper
choices of the initial $f$, the above condition can be written as 
\be 
\zt \ge
\frac{2 t_0}{\zg - \zd} \cdot \frac{J^{*} F_e V}{k_B T} ~, \label{tauconditionJ} 
\ee
where $J^{*}$ is the maximum
of the flux. This shows that, differently from the case of the steady state
$\zL$-FR, the convergence time of the $\zW_\infty$-FR does not increase when 
the dissipative field tends to 0; to the contrary, it may decrease. 

As in the transient case, an asymptotic $\phi$-FR holds for $A$
in a range which depends on the dynamics and on the observable,
but for large $\zt$ it remains close to $(-\zW^{*},\zW^{*})$. 
Indeed, one has
\be 
-\frac{2 t + \zt}{\zt} \zW^{*} \le \frac{1}{\zt}
\ln \frac{\mu_{t}(\overline{\phi}_{0,\zt}|_{A^+_\zd})}
{\mu_{t}(\overline{\phi}_{0,\zt}|_{A^-_\zd})} \le
\frac{2 t + \zt}{\zt} \zW^{*} 
\ee
and taking any $\zg>0$
and letting $t$ grow, with $\zt(t) \gg t$, one can write 

\vskip 10pt \noi
\textbf{$\phi_\infty$-FR:}
\be
- \zW^* - \zg
\le  \lim_{t \to \infty}\frac{1}{\zt(t)} \ln
\frac{\mu_{t}(\overline{\phi}_{0,\zt(t)}|_{A^+_\zd}) }
{\mu_{t}(\overline{\phi}_{0,\zt(t)}|_{A^-_\zd})} \le 
\zW^* + \zg ~,
\label{aribond}
\ee
for any odd $\phi$, and
with the smallest $\zW^{*}$, among the upper bounds of the allowed
dissipation functions.

\vskip 2pt \noi
Because at equilibrium the 
dissipation function given by Eq.(\ref{epr}) is $\zW=-F_{e}JV/k_{_{B}}T=0$, one has
\be 
-\zg \le \frac{1}{\zt} \ln \frac{\mu_{t_0}(\overline{\phi}_{0,\zt}|_{A^+_\zd})} {\mu_{t_0}(\overline{\phi}_{0,\zt}|_{A^-_\zd})} \le \zg ~, \label{equilB} 
\ee
whatever odd variable
$\phi$ is considered, $\zW$, $\zL$ or $J$ for instance. This yields the usual equilibrium
symmetry, in the full range of $J$, not just in 0, and without the need
for long convergence times.

Given  the dynamics $S^{\zt}$, either there is a unique allowed 
density $f_0$, hence a unique dissipation function $\zW$ or, given a 
second initial density $\tilde{f}_0$, there is a corresponding 
$\tilde{\zW} \ne \zW$. If this is the case, and the attractor for 
both $f_0$ and $\tilde{f}_0$ is the same, one can write
\be 
\left\langle \exp \left( -\tilde{\Omega}_{0,t} \right)
\right\rangle_{\Ft \in A^+_\zd} \approx
\frac{\mu_{t_0}(\overline{\phi}_{0,\zt}|_{A^-_\zd})} 
{\mu_{t_0}(\overline{\phi}_{0,\zt}|_{A^+_\zd})} \approx 
\langle \exp \left( -\Omega_{0,t} \right) \rangle_{\Ft \in A^+_\zd} ~, 
\label{corresp}
\ee
for large $t_{0}$, all $\zt$ and all allowed $A$. Then, as far as the 
asymptotic fluctuation relations are concerned, the different dissipation 
functions are equivalent. 

Let us note that Eq.(\ref{almostSS}) does not necessarily imply the steady
state $\zW$-FR, even when there is a single steady state and all initial measures
$\mu_0$ with density $f_0$ are attracted by the same $\mu_{\infty}$, because
the limit in Eq.(\ref{aribond}) could be singular. 
Although this is not impossible, it is not 
the typical situation of particle systems of physical interest (cf.\ Section 4
and Appendix B), and
Eq.(\ref{almostSS}) is a result rather close to the steady state $\zW$-FR.

In Section 4 we discuss the conditions under which Eq.(\ref{almostSS}) describes the statistics of trajectory segments sampled from a single steady state trajectory.  
Before we do this, we note that 
whereas in Sections 2 and 3, the initial distribution $f_0$ can have arbitrary form, 
provided it is ergodically consistent with the dynamics considered, a further condition on $f_0$ is imposed, in order to obtain the steady state $\zW$-FR: the decay of the $\zW$-autocorrelation,
with respect to $f_0$. This is necessary for the initial state to relax to a steady state. 
The further requirement that the ensemble is not only invariant under $S^t_0$, but that is 
also generated by a single trajectory, has the purpose to avoid the presence of distinct
steady states, selected by different choices of the initial conditions. These
states, indeed, would have different properties, and a unique $\mu_\infty$ would 
not be possible.
For instance, if $f_0$ was selected to be a uniform distribution (so that 
$\Omega=-\Lambda$) and isokinetic dynamics was carried out, a distinct $\mu_\infty$
would be expected for each distinct value of the kinetic energy.

If the system under investigation does not converge to a steady state with 
positive and negative fluctuations, there is no need for a steady state relation, 
but the $\zW$-FR, the $\phi$-FR, the $\zW_\infty$-FR and the $\phi_\infty$-FR 
remain valid for the given ensemble of trajectories.

\section{STEADY STATE FLUCTUATION RELATIONS}
We assume the steady state exists and that it is unique. This will not always be the case, for example if we start from a canonical phase space distribution at time zero and have isokinetic dynamics subsequently, there will be distinct 
steady state attractors for each trajectory with a different 
kinetic energy. Each of the steady state attractors will have different properties because they are at different kinetic temperatures.  By saying that the steady state is unique we mean that, apart from a set of measure zero, the statistics generated by each 
trajectory is independent of the initial 
phase, in the relevant equilibrium phase space.  

Equations (\ref{exactFT_t_0}) and (\ref{exactFTt_01}) are both exact, but contain 
information on the trajectories before the steady state is reached.  
To be of practical use in the description of the steady state fluctuations of the 
dissipation function, the last two terms in the equation have to be negligible. 
Therefore, for $A \ne 0$, some assumption on the nature of the logarithmic term of 
Eq.(\ref{exactFTt_01}) is necessary since, in
principle, this term could be of order $O(t_{0}/\zt)$, and $t_{0}$ should tend 
to infinity, in order to reach the steady state, before $\zt$ does \cite{CG99}. 
We argue that, in most cases of interest, the dynamics of particle systems 
used to model nonequilibrium fluids satisfy the following:

\vskip 10pt \noi
{\bf Steady state $\zW$-FR.}
{\it For any {\em tolerance} $\zg>0$, there exists $\zd_\zg , \zt_\zg > 0$
such that 
\be 
A - \zg \le \frac{1}{\zt}
\ln \frac{\mu_{\infty}(\overline{\zW}_{0,\zt}|_{A^+_\zd})}
{\mu_{\infty}(\overline{\zW}_{0,\zt}|_{A^-_\zd})} \le A + \zg 
\label{SSFTestim} 
\ee
holds for $0 < \zd \le \zd_\zg$ and $\zt \ge \zt_\zg$, if both the values 
$A$ and $-A$ are {\em $\zd$-possible}.}

\vskip 10pt \noi 
Here, the following definition has been used:

\vskip 10pt \noi      
{\bf Definition.} {\it
The value $A \in \zR$ is called {\em $\zd$-possible} if, given $\zd>0$, there 
exists $\zt_{A,\zd}>0$ such that 
$$
\mu_{\infty}(\overline{\zW}_{0,\zt}|_{A_{\zd}^{+}})>0
$$ 
for all $\zt>\zt_{A,\zd}$}.\footnote{Because of the finiteness of $\zt$, 
this can be the case even though
$\mu_{\infty}(\overline{\zW}_{0,\zt}|_{A^+_\zd})\to0$, for $\zt \to \infty$,
as it does if the steady state ensemble average of $\zW$, 
$\langle \zW \rangle_\infty$ say, does not belong to $A^+_\zd$.} 

\vskip 10pt \noi
It follows that, for $A$ to be $\zd$-possible,  
$\mu_{\infty}(\overline{\zW}_{0,\zt}|_{A_{\zd}^{+}})$ must decay with $\zt$ more 
slowly than the error term
$\left[ \ze(\zd,t_0,A,\zt) - \frac{1}{\zt} \ln \left\langle e^{-\zW_{0,t_0}
- \zW_{t_0+\zt,2t_0+\zt}} \right\rangle_{\Wt \in A^+_\zd} \right]$ of 
Eq.(\ref{exactFTt_01}). Indeed,
given $\zd > 0$, a pair $(-A,A)$ and a fixed $\zt$, the logarithmic term 
in the right hand side of Eq.(\ref{exactFTt_01}) may either take arbitrarily 
large values or be bounded, in the $t_0 \to \infty$ limit. In the first 
case, because $A, \zd$ and $\zt$ are fixed real numbers and $|\ze| \le \zd$, 
the corresponding divergence is reflected  on the measures 
$\mu_{t_0}(\overline{\zW}_{0,\zt}|_{A^+_\zd}))$ and
$\mu_{t_0}(\overline{\zW}_{0,\zt}|_{A_{\zd}^-})$, one of which at 
least either tends to zero or does not converge at all. Indeed, if for 
every $R,\zt,\hat{t}_0>0$ there is $t_0 > \hat{t}_0$ such that
\be
R < \frac{\mu_{t_0}(\overline{\zW}_{0,\zt}|_{A^+_\zd})}
{\mu_{t_0}(\overline{\zW}_{0,\zt}|_{A^-_\zd})} ~,
\label{faster}
\ee
one has 
\be
R < \frac{1}{\mu_{t_0}(\overline{\zW}_{0,\zt}|_{A^-_\zd})} ~, \qquad \mbox{or } 
~~~~
\mu_{t_0}(\overline{\zW}_{0,\zt}|_{A^-_\zd}) < \frac{1}{R}
\label{faster1}
\ee
which corresponds either to 
$\mu_{\infty}(\overline{\zW}_{0,\zt}|_{A^-_\zd})=0$, or to the 
$\mu_{\infty}$ non-measurability of $\overline{\zW}_{0,\zt}|_{A^-_\zd}$. 
In either case, $-A$ is not $\zd$-possible, and similarly one may treat $A$. 

Clearly, $A$ is of no interest for a steady state FR, i.e.\ it does not 
belong to the domain of the steady state $\zW$-FR with tolerance $\zg$, if
$A$ and $-A$ are not both $\zd$-possible; in a number of cases that domain 
may even be empty.\footnote{There are various reasons for which $A$ or $-A$ might 
not be in the the domain of the steady state $\zW$-FR. For example, $A$ could be 
larger than the largest value of $\zW$, or there could be no fluctuations in 
the steady state (cf.\ Appendix B).} 
Therefore, we focus only on the $A$ values with bounded logarithmic 
term in Eq.(\ref{exactFTt_01}), and let $M(A,\zd,\zt)$ be the real number 
to which the conditional average in Eq.(\ref{exactFTt_01}) tends,
in the $t_0 \to \infty$ limit. We may then write
\be
A - \frac 1 \zt \ln M(A,\zd,\zt) - \zd \le \frac{1}{\zt}
\ln \frac{\mu_{\infty}(\overline{\zW}_{0,\zt}|_{A^+_\zd})}
{\mu_{\infty}(\overline{\zW}_{0,\zt}|_{A^-_\zd})} \le A - 
\frac 1 \zt \ln M(A,\zd,\zt) + \zd ~.
\label{step1}
\ee
If $(1/\zt)\ln M(A,\zd,\zt)$ tends to zero, for the given $A,\zd$ and
growing $\zt$, the steady state $\zW$-FR holds for $A$. This is the case 
if correlations decay, so that the thermodynamic behavior sets in. In 
particular, if the $\zW$-autocorrelation function is a $\delta$-function
(as e.g.\ in Ref.\cite{BGG,BGGZ}), one can write:
\be
\left\langle e^{-\zW_{0,t_0} - \zW_{t_0+\zt,2t_0+\zt}} 
\right\rangle_{\Wt\in A^+_\zd} = 
\left\langle e^{-\zW_{0,t_0} - \zW_{t_0+\zt,2t_0+\zt}} \right\rangle =
\left\langle e^{-\zW_{0,t_0}} 
\right\rangle \left\langle e^{- \zW_{t_0+\zt,2t_0+\zt}} 
\right\rangle ~,
\label{condave-full}
\ee
and 
\be
1=\left\langle e^{-\zW_{0,s} -\zW_{s,t} } \right\rangle = 
\left\langle e^{-\zW_{0,s}} \right\rangle \left\langle e^{-\zW_{s,t}} 
\right\rangle ~, 
\label{immedia}
\ee
which implies
\be
\left\langle e^{-\zW_{s,t}} \right\rangle =1 ~~~~ \mbox{for all ~} s,t
\label{implies}
\ee
because of the NonEquilibrium Partition Identity (\ref{interN}).
Then, $(1/\zt)\ln M(A,\zd,\zt)$ identically vanishes for all $\zt$.
This is an idealized situation and, 
indeed, the tests performed on molecular dynamics systems,\footnote{The
conditional average of Eq.(\ref{exactFTt_01}) can, of course, be computed 
independently of the other terms in Eq.(\ref{exactFTt_01}), although this
is not strictly necessary, because Eq.(\ref{exactFTt_01}) is an exact identity.} 
cf.\ Ref.\cite{ESRnumer}, indicate that 
typically there exists a constant $K$ (often not too far from 1), such that 
\be
0 < \frac{1}{K} \le \left\langle e^{-\zW_{0,t_0} - \zW_{t_0+\zt,2t_0+\zt}} 
\right\rangle_{\Wt \in A^+_\zd} \le K ~,
\label{aveBDD}
\ee
which may be understood as follows. In the first place, the conditional 
average is limited for growing $t_0$, if $-A,A$ are $\zd$-possible, and 
equals 1 for $\zt=0$. Then, if the $\zW$-autocorrelation decays with 
time scale $t_M$,\footnote{This time is called Maxwell time; its order 
of magnitude is that of the mean free time. The Maxwell time expresses 
a physical property of the system and depends only mildly on the external 
field; usually, $t_M(F_e) = t_M(0) + O (F_e^2)$} 
one expects $\ln K$ to be of order $O(t_M)$.
Taking $\zt$ sufficiently larger than $t_M$, and letting $c_A$ be a natural
number dependent on the given $A$, one may write
\bea
\hskip -20pt
&&\hskip -70pt \left\langle
e^{-\zW_{0,t_0} - \zW_{t_0+\zt,2t_0+\zt}} 
\right\rangle_{\overline{\zW}_{t_0,t_0+\zt} \in A_\zd^+} = \nonumber \\
&=&
\left\langle e^{-\zW_{0,t_0-c_A t_M} - \zW_{t_0+\zt+c_A t_M,2t_0+\zt}}
\cdot e^{-\zW_{t_0-c_A t_M,t_0} - \zW_{t_0+\zt,t_0+\zt+c_A t_M}}
\right\rangle_{\overline{\zW}_{t_0,t_0+\zt} \in A_\zd^+}  \nonumber \\
&\approx& \left\langle e^{-\zW_{0,t_0-c_A t_M}} \cdot e^{- \zW_{t_0+\zt+c_a t_M,2t_0+\zt}}
\right\rangle_{\overline{\zW}_{t_0,t_0+\zt} \in A_\zd^+} \label{newargdeco} \\
&\approx& \left\langle e^{-\zW_{0,t_0-c_A t_M}} \cdot e^{- \zW_{t_0+\zt+c_a t_M,2t_0+\zt}}
\right\rangle \nonumber  \nonumber \\
&\approx& \left\langle e^{-\zW_{0,t_0-c_A t_M}} \right\rangle
\left\langle e^{- \zW_{t_0+\zt+c_A t_M,2t_0+\zt}} \right\rangle 
= \left\langle e^{- \zW_{t_0+\zt+c_A t_M,2t_0+\zt}} \right\rangle  
\nonumber
\eea 
with an accuracy which improves with increasing $t_0$ and $\zt$, 
if $c_A$ is the number of Maxwell times necessary to obtain the desired
decorrelation at the chosen $A$. Indeed, one would eventually obtain
$t_0 , \zt \gg c_A t_M$.

If Eq.(\ref{newargdeco}) holds, the $t_0 \to \infty$ limit of 
$\left\langle e^{- \zW_{t_0+\zt+c_A t_M,2t_0+\zt}} \right\rangle$ is finite for all 
$\zd$-possible pairs $(-A,A)$, and is insensitive to variations of $\zt$ above a 
certain threshold. Indeed, let $\hat{\zt}=\zt+c_A t_M$ and observe that
\be
\left\langle e^{- \zW_{t_0+\hat{\zt},2t_0+\zt}} \right\rangle = \int 
e^{- \zW_{t_0+\hat{\zt},2t_0+\zt}(\zG)} f_0(\zG) d \zG
= \int e^{- \zW_{0,t_0-c_A t_M}(\zG)} f_{t_0+\hat{\zt}}(\zG) d \zG := 
\left\langle e^{- \zW_{0,t_0-c_A t_M}} \right\rangle_{t_0+\hat{\zt}}
\label{shiftedave}
\ee
is the average of $e^{- \zW_{0,t_0-c_A t_M}}$ computed with the probability 
distribution obtained from the evolution of the initial one, for the 
time $t=t_0+\hat{\zt}$. The value of this average may be estimated noting 
that, for $\zd$-possible pairs $(-A,A)$, and growing $t_0$, it converges 
approximately to the positive number $M(A,\zd,\hat{\zt})$, if 
Eq.(\ref{newargdeco}) holds. Furthermore, for fixed $t_0>0$ and for
$t$ of the order of the Maxwell time, 
$\left\langle e^{- \zW_{0,t_0-c_A t_M}} \right\rangle_{t}$ approaches its 
steady state mean value, $\left\langle e^{- \zW_{0,t_0-c_A t_M}} \right\rangle_\infty$ which is equal to 1 at equilibrium for all $t_0$ and $c_A$.
Then, for sufficiently large $t_0$ and $t=t_0+\hat{\zt}$, one may write
\be
M(A,\zd,\hat\zt) \approx \left\langle
e^{-\zW_{0,t_0} - \zW_{t_0+\zt,2t_0+\zt}}
\right\rangle_{\overline{\zW}_{t_0,t_0+\zt} \in A_\zd^+} \approx
\left\langle e^{- \zW_{t_0+\hat{\zt},2t_0+\zt}} \right\rangle =
\left\langle e^{- \zW_{0,t_0-c_A t_M}} \right\rangle_{t}
\approx \left\langle e^{- \zW_{0,t_0-c_A t_M}} \right\rangle_\infty
\label{limits}
\ee
thanks also to the fact that $\hat\zt$ is
by itself larger than the decorrelation time. Because the left hand side of 
Eq.(\ref{limits}) does not depend on $t_0$ while the right hand side does not
depend on $\zt$, one concludes that further growths of $t_0$ or of $\zt$ do not 
cause any substantial changes in the conditional average in the middle. Close to
equilibrium, this term should only have a correction of order $O(F_e^2)$ to its
equilibrium average, 1.
Numerical studies show that this is precisely what happens 
to the quantities in Eqs.(\ref{aveBDD},\ref{newargdeco},\ref{shiftedave}),
\cite{ESRnumer}.

Note that $c_A$ may vary with $A$, as the set of trajectories involved 
in the conditional average changes with $A$; $c_A$ could then even diverge.
However, close to equilibrium at least, the existence of the transport 
coefficients implies that the conditional correlation functions must 
not grow too much (cf.\ Appendix B). 

Note also that, for the steady state \WFR to hold, it suffices that 
$(1/\zt) \ln M(A,\zd,\zt)$ can be made as small as one wishes, by taking 
sufficiently large $\zt$ and small $\zd$. Thus, it is not necessary that 
Eq.(\ref{aveBDD}) holds; it suffices that $M(A,\zd,\zt)$ grows 
less than exponentially fast with $\zt$, or even that it grows exponentially 
fast, with a rate which tends to zero when $\zd$ does, as in Eq.(\ref{1stprop}),
where $A=0$. 
However, the numerical tests performed so far indicate that Eq.(\ref{aveBDD}) is 
usually satisfied in the full range of numerically or experimentally accessible
$A$'s. 

If the $\zW$-autocorrelation decays, and Eq.(\ref{aveBDD}) holds with the 
condition on $\zW$ replaced by an equivalent one for another observable 
$\phi$, one obtains the

\vskip 10pt\noi
{\bf Steady State $\phi$-FR.} {\it For sufficiently large $\zt$,}
\be
\lim_{t_0 \to \infty}
\frac{\ln \langle e^{-\Omega_{t_0,t_0+\zt}}
\rangle^{-1}_{\Ft \in A^+_\zd}}{\zt} - \zg \le
\frac 1 \zt \ln \frac{\mu_{\infty}(\overline{\phi}_{0,\zt}|_{A^+_\zd})}
{\mu_{\infty}(\overline{\phi}_{0,\zt}|_{A^-_\zd})} \le  
\lim_{t_0 \to \infty} \frac{
\ln \langle e^{-\Omega_{t_0,t_0+\zt}}
\rangle^{-1}_{\Ft \in A^+_\zd}}{\zt} + \zg
\ee

\vskip 10pt\noi
This expression states that the
ratio of the probabilities that $\overline{\phi}_{0,\zt} \in A^+_\zd$ and 
$\overline{\phi}_{0,\zt} \in A^-_\zd$ is determined by the average of the
exponential of the values taken by $\zW$ in the same time intervals in which
$\phi$ has average in $A_\zd^+$. Because $\phi$ can be any odd observable,
$\zW$ appears to be a rather special function of phase.

One may wonder how the above could be reconciled with the modifications of 
the steady state $\zL$-FR suggested for a special kind of non-transitive 
dynamical systems \cite{CG99,axiomC,BGG}. As explained in Appendix B, 
Eq.(\ref{aveBDD}) is not in contradiction to the scenarios of Refs.\cite{CG99,axiomC,BGG}, 
because the decay of correlations enjoyed by their dynamical systems concern the invariant 
measures, while the conditional average in Eq.(\ref{aveBDD}) is computed with respect to the
initial measure. Therefore, 
the arguments of this section and those of Refs.\cite{axiomC,BGG} 
are compatible, since Eq.(\ref{aveBDD}) would not hold if the cases of Refs.\cite{CG99,axiomC,BGG} were realized. 
The scenarios under which the approximation (\ref{aveBDD}) 
does not hold, appear however quite peculiar, and this section explains 
why quite commonly the steady state \WFR does hold: time reversibility and 
the decay of the $\zW$-autocorrelation characterize the most common 
deterministic models of nonequilibrium statistical mechanics. 
Through the connection with the Maxwell time $t_M$, this approach also justifies why the 
convergence times of the steady state \WFR do not grow
with decreasing dissipation: the fact is that the \WFR
depends only on material properties of the system, and 
these do not change much around equilibrium. This is 
quite different from the case of the $\zL$-FR, whose 
convergence times increase as the driving is decreased, 
and typically diverge in the equilibrium limit. As stated above,
the calculations of this section hold for any regular density $f$, i.e.\ for all 
possible dissipation functions $\zW$, but the \WFR turns into the $\zS$-FR only 
when the initial distribution can be generated by single field free trajectories,
and $AI\zG$ holds.

\section{CONCLUSIONS}
In this paper we have investigated the mechanisms underlying the validity
of the transient and steady state FR's, following the 
prescriptions given by Evans and Searles over the last decade. As already 
evidenced in the literature, the transient relations hold under very general
conditions; it suffices that the dynamics of the particle systems
at hand be deterministic, autonomous, and time reversal invariant. If the
initial state is properly selected, the corresponding $\zW$ equals the 
energy dissipation rate $\zS$.
Different initial densities $f$ are allowed, for any given dynamics
$S^{\zt}$; they are not completely equivalent, as in general they
generate different dissipation functions $\zW$, but to some extent
they are interchangeable (cf.\ Eq.(\ref{corresp})).
The transient relations describe the properties of evolving ensembles of 
systems, even in their asymptotic form.

The steady state fluctuation relations need further assumptions. Firstly
and quite obviously the system must relax to a nonequilibrium steady state. 
This requirement is nontrivial. The necessary and sufficient conditions
for this to occur are unknown, but necessarily involve some 
loss of memory of the initial state, and the corresponding decay of
correlations. Certainly there are examples of dissipative
particle systems in contact with a thermal heat bath that do not relax to 
a steady state. 
Also, there may be distinct basins of attraction in ${\cal M}$ 
with distinct steady state measures. For example if ${\cal M}$ is taken to be 
all of phase space with an energy less than some bound and the initial ensemble
is canonical, but the dynamics is isoenergetic, then obviously there will be no
mixing between trajectories that have different initial energies, and the
subensemble members will generate distinctly different energy dependent steady
states. Different steady states may coexist also when dissipation is too high
\cite{GRV}. Transitivity (i.e.\ that a typical trajectory explores the whole
phase space relevant to the given dynamics) avoids these difficulties. 

As a matter of fact, most of the molecular dynamics studies performed so far 
indicate that convergence to a steady state independent of the initial phase 
is a very common situation. This convergence is consistent with laboratory 
experiments where for systems with low Reynolds/Raleigh numbers the nonequilibrium 
steady state (if such a state exists) is quite independent of the initial starting 
state or the preparative history of the system. The low shear rate viscosity of 
fluids is empirically observed to be a function of the temperature and density 
(or pressure) of the fluid. The independence of the initial phase or preparative 
history is the nonequilibrium analogue of the situation where at equilibrium 
thermodynamic free energies are commonly observed to be thermodynamic state functions. 
Even at equilibrium the conditions required for this to be the case are not known 
(they are violated in glassy systems for example).

If a single steady state is approached, its statistics may be recovered from an 
individual trajectory, and this may have no knowledge of the initial 
ensemble $f$. This loss of knowledge of the 
initial ensemble amounts to a decay of correlations, 
which may be further weakened. It must be stressed 
that the decay of correlations required by the validity of the \WFR reduces
to the $\zW$-autocorrelation decay, with respect to the initial 
ensemble $f$. As is well known, this is required, close to equilibrium,
for the existence of the transport coefficients. 

In the case that transitivity and boundedness of $\zW$ 
suffice to obtain the steady state $\zW$-FR (something that 
needs further study), we observe that these are much weaker conditions than 
the Anosov assumption, which also implies the boundedness of the phase space
contraction rate.%
\footnote{Recall that the Anosov assumption was considered appropriate for the
particle system of \cite{ECM2}, in which $-\zL=\zW$, although $\zW$
was not bounded.%
} This is interesting, because one can name a number of cases which do have bounded $\zW$,
like the isokinetic (or isoenergetic) hard particle systems; the isokinetic
or Nos\'{e}-Hoover thermostatted systems for heat current and shear
flow systems with bounded interaction potentials; the color conductivity
isokinetic systems with unbounded potentials etc. Transitivity, in turn, 
is almost always verified.

We note that in order to obtain a steady state $\zS$-FR, the initial 
distribution, $f_0$, must be obtainable from a single the field free evolution, 
and $AI\zG$ must hold. These conditions are easily obtainable for systems
of physical interest.

The $\zS$-FR's directly refer to the dissipative fluxes or to quantities simply 
related to them, rather than to the phase space contraction rate. This avoids 
the difficulty of justifying how a $\zL$-FR can speak of the fluctuations
of the entropy production rate. As discussed in \cite{ESR}, these
difficulties are more evident close to equilibrium, where one observes
that the convergence times of the $\zL$-FR typically diverge as the 
dissipative applied field decreases
and the system approaches equilibrium. This is due to the fact that
$\zL$ contains undesirable terms, which have to average to zero.
As one approaches equilibrium the dissipation 
decreases as the square of the dissipative field while the undesirable
terms remain constant (to leading order in the field). Therefore,
longer and longer averaging times are required before the $\zL$-FR 
converges. It has been argued \cite{romans} that the $\zS$-FR can be obtained 
from the $\zL$-FR because the time averages of $\zS$ and $-\Lambda$ become 
equivalent in the long time limit. However, this argument can not explain the 
satisfaction of the $\zS$-FR at physically accessible times in non-isoenergetic 
systems at low fields. Moreover, if the Van Zon - Cohen picture \cite{othertheory}
applies to deterministic particle systems, as argued in Refs.\cite{ESR,BGGZ,TG}, 
the asymptotic equivalence between $\zS$-FR and $\zL$-FR is restricted to a limited 
fraction of the possible fluctuation values. As a matter of
fact, $\zS$ equals $-\zL$ only if $f$ is uniform in $\mathcal{M}$, which 
requires $\mathcal{M}$ to be compact. When this is not the case, $\zS$
and $\zL$ are substantially different quantities, as even Ref.\cite{BGGZ} 
shows. The difficulties with the weak 
field convergence of $\zL$-FR make the derivation of Green-Kubo relations from 
$\zL$-FR, highly problematic \cite{ESR}. 
Differently, if a relation for the dissipation is obtained independently, 
this difficulty is not present. In particular, the convergence times 
are understood in terms of material properties related
to known, physical properties of the system (e.g. the Maxwell time
is related to the infinite frequency shear modulus and the zero shear
rate shear viscosity), and the derivation of the Green-Kubo
relations from the steady state $\zS$-FRs is reasonably straightforward \cite{ESR}. 

It is also important to consider the consistency of the results obtained in this manuscript with those obtained for the axiom C systems postulated in Refs.\cite{axiomC,BGG}.  As discussed in Appendix B, these systems may satisfy
the decay of correlations of all observables, with respect to an invariant
measure supported on a surface $\mathcal{M}_r$, of dimension lower than that 
of the phase space $\mathcal{M}$. Then, assuming as in Ref.\cite{BGG}, 
that $\zL$ is proportional to the contraction rate of volumes in $\mathcal{M}_r$, $\zL_r$
say, $\zL$ should obey a modified FR.
So far there is no direct evidence, in particle systems, of the validity of
that assumption (supported by numerical studies of certain
hydrodynamic models \cite{GG-GNS}). In fact, in the case considered in Ref.\cite{stephennew}, the dissipation is sufficiently high that the dynamics
is not transitive yet the $\zS$-FR holds without the modifications conjectured in \cite{BGG}. 
In other words, our work and Refs.\cite{axiomC,BGG} usually consider different 
observables, and when they consider the same observable, it is not necessary 
that they lead to the same conclusions of Refs.\cite{axiomC,BGG}. If, on the 
other hand, the treatment of Refs.\cite{axiomC,BGG} applies, 
the decay of correlations of Section 4 is compatible with it,
as explained in Appendix B, because it is referred to the initial measure,
and not the invariant one.\footnote{One should also observe that 
Eq.(\ref{exactFTt_01}) is exact, hence cannot lead to any contradiction 
about reversible dynamical systems.}
The energy dissipation $\zS$ remains related to $\zL$, not to $\zL_r$, even if the steady state 
is confined inside a lower dimensional manifold $\mathcal{M}_r$.

As pointed out by one of the anonymous referees, 
the existence of a unique stationary measure may be considered a strong assumption, 
and the necessary and sufficient condition required for it to hold are not known. 
Therefore, the study of Anosov systems constitutes a natural starting point for the 
purpose of classifying the dynamical systems that verify some kind of FR. The 
purpose of the present paper is different: it amounts to analyzing systems which 
cannot be chosen at will, but are selected by physical conditions, in order to
understand further their properties as well as the physical mechanism responsible for 
FR's to hold in natural systems. We have found that time 
reversibility and ergodic consistency imply, without further assumptions, a number 
of transient and asymptotic FRs (cf.\ Sections 2 and 3), 
which are amenable to experimental verification. If the system of interest does 
not reach a unique steady state, assuming that it is Anosov does not help of course, 
but those relations remain valid. 
When present, convergence to a unique steady state is the manifestation of a certain 
decay of correlations of the observables, which typically are required to relax to a steady state. 
The approach proposed in this paper thus makes it clear that
the details of the microscopic dynamics (like the very strong hypothesis
of uniform hyperbolicity) do not play any role in the validity of the
FR's, as appropriate for thermodynamic relations. The class of dynamical
systems which obey Eq.(\ref{aveBDD})
cannot be determined, however, by this approach.

These facts may help in developing further the theory of 
nonequilibrium phenomena, as the derivation of the novel relations given above indicates.

\section*{ACKNOWLEDGMENTS}
We wish to thank the Australian Research Council, and the Italian Embassy in 
Canberra for support. We also thank the Erwin Schr\"{o}dinger Institute for 
support for the "Workshop on Stochastic and Deterministic Dynamics in 
Equilibrium and Nonequilibrium Systems" where this work was discussed.  
We thank Barbara Johnston for her input to this work, and Stephen Williams, 
Emil Mittag, Angelo Vulpiani, E.G.D. Cohen, Federico Bonetto and Carlos Mejia Monasterio for their 
comments. Thanks are in order also to the anonymous referees, for useful
remarks on the purpose of our work.

\newpage{} 

\section*{APPENDIX A}

Using the general definition of the dissipation function (\ref{omegat}),
it can be shown that the instantaneous value of the \textit{dissipation
function} is directly related to the instantaneous \textit{dissipative
flux} for a wide class of systems. Here we consider three cases: isoenergetic,
isokinetic and Nos\'{e}-Hoover thermostatted dynamics.

A general form of the equations of motion for a field-driven \textit{N}-particle,
thermostatted nonequilibrium system is 
\bea 
\qdot&=&\textbf{p}_i/m+\textbf{C}_i(\zG)\cdot\textbf{F}_e \nonumber \\
 \pdot&=&\textbf{F}_i+\textbf{D}_i(\zG)\cdot\textbf{F}_e-\za\textbf{p}_i
\label{geneqnsmotion} 
\eea 
where ${\bf F}_{e}$ is the applied field
that is coupled to the system via the phase functions ${\bf C}_{i}(\zG)$
and ${\bf D}_{i}(\zG)$. The term $-\za{\bf p}_{i}$ is 
a deterministic, time reversible term used to add or remove
heat from the system \cite{DenisGarybook}. The adiabatic
equations of motion 
\bea 
\qdot&=&\textbf{p}_i/m+\textbf{C}_i(\zG)\cdot\textbf{F}_e \nonumber \\
 \pdot&=&\textbf{F}_i+\textbf{D}_i(\zG)\cdot\textbf{F}_e \label{adiabeqnsmotion}
\eea 
simply lack the thermostatting term. If Eqs.(\ref{adiabeqnsmotion})
are Hamiltonian, the phase space expansion rate $\zL$ obviously vanishes,
a condition referred to as \textit{adiabatic incompressibility of
phase space}, $AI\zG$ \cite{DenisGarybook}, but $AI\zG$ may hold
even if the general adiabatic equations are not Hamiltonian. Here, for simplicity,
we concentrate on systems which satisfy $AI\zG$, because their dissipative
flux and dissipation function are simply related. In the other cases,
the relationship between dissipative flux and dissipation function
is only more complicated. We also assume that the initial 
distribution of phases $f_0$ may in principle be generated by a single
field-free ($F_e=0$) thermostatted dynamics. By this we mean that $f_0$ must 
not only be preserved by the equilibrium dynamics, but also that a single 
field free trajectory may explore all of its support. In other words, that 
this support is not the union of disjoint invariant sets, like the constant 
energy surfaces corresponding to different initial energies of Hamiltonian 
systems. 
This is consistent with nonlinear response theory that shows that if one is interested in the statistics of the physically 
relevant steady state observables, $f_0$ should be that which corresponds to the field free
dynamics \cite{DenisGarybook,ESnew}.  The possible difficulties that might be encountered if $f_0$ is not selected to be the distribution function generated by the field free dynamics can easily be seen by considering the distribution of                                                  $\overline{\phi}_{0,\zt}$ along an equilibrium trajectory.  In this
case the correct physical picture requires the right hand side of
Eq.(\ref{ESSFT}) to be 1 for all $A, t_0, \delta$
and $\tau$.  This can only be the case if $\Omega(\Gamma)=0$ for all $\Gamma$.
Using Eq.(\ref{ft_0WOmega}),
this implies that $f_0(S^t_0\Gamma)=f_t(S^t_0\Gamma)$ under the field free dynamics $S^t_0$,
and that $f_0=f_t$ must be preserved by the field free dynamics.
Note that a single trajectory can't possibly generate the ensemble in physically 
relevant times, in any system made of more than a few particles, but it may do
it, in principle, allowing it to evolve for exceedingly long. This is
needed for the possible initial equilibrium state to be uniquely determined.

In general, the dissipative flux, ${\bf J}$, is obtained from the
adiabatic time-derivative of the internal energy, $H_{0}=\sum_{i=1}^{N}\frac{{\bf p}_{i}\cdot{\bf p}_{i}}{2m}+\Phi({\bf q})$,
i.e. from the derivative of the internal energy under the dynamics
of Eqs.(\ref{adiabeqnsmotion}):

\be
{\dot H}^{ad}_0=-\sum_{i=1}^N(\frac{{\bf p}_i}{m}\cdot{\bf D}_i-{\bf F}_i\cdot{\bf C}_i)\cdot {\bf F}_e
:= - {\bf J}V\cdot{\bf F}_e~,
\label{H0dotad}
\ee
while the full dynamics of Eqs.(\ref{geneqnsmotion}) yields

\be 
{\dot H}_0=- \textbf{J}V\cdot\textbf{F}_e-\za \sum_{i=1}^N
\frac{\textbf{p}_i\cdot\textbf{p}_i}{m}=- \textbf{J}V\cdot\textbf{F}_e-2K\za
\label{H0dot} 
\ee
were $K$ is the kinetic energy of the system.
Consider now three widely used deterministic, reversible thermostatting
methods.

\begin{itemize}
\item Gaussian isoenergetic (ergostatted)
system. In this case $H_{0}$ is fixed, i.e.\ $\dot{H}_{0}=0$,
and Eq.(\ref{H0dot}) gives 
\be 
\za(\zG)=- \frac{\textbf{J}(\zG)V\cdot\textbf{F}_e}{2K(\zG)}
\label{alphaJ} 
\ee

For the ergostatted system described
by equations (\ref{geneqnsmotion}), the phase space expansion rate
is, 
\be 
\zL(\zG) = \nabla \cdot \dot{\Gamma} =\sum_{i=1}^N
\left(\frac{\partial \textbf{C}_i(\zG)}{\partial \textbf{q}_i}+\frac{\partial
\textbf{D}_i(\zG)}{\partial \textbf{p}_i}\right)\cdot \textbf{F}_e-dN\za(\zG)+O_N(1)
\stackrel{AI\zG}{=} -dN\za(\zG)+O_N(1) \label{LambdaE0} 
\ee
where $AI\zG$ is used to obtain the final equality, and $O_{N}(1)$
is a correction which is order 1 in $N$. The $O_{N}(1)$ term can
be explicitly determined from the partial derivative of the Gaussian
ergostatting term with respect to ${\bf p}$. The equilibrium phase
space distribution function is microcanonical on a surface $H_{0}=$const,
hence $f(iS^{\zt}\Gamma)=f(S^{\zt}\Gamma)=f(\Gamma)$, and Eq.(\ref{omegat})
yields: 
\be 
\overline{\Omega}_{0,\zt}^{isoE}(\zG)=-\overline{\Lambda}_{0,\zt}(\zG)
\ee
for all $\zt$, so 
\be 
\zW(\zG)=-\zL(\zG)=dN\za(\zG)+O_N(1)=-\textbf{J}(\zG)\cdot\textbf{F}_e
\beta (\zG) V + O_N(1) \label{OmegaJE0} 
\ee
where $
\beta(\zG)=(dN-d-1)/(2K(\zG))$.
\item Gaussian isokinetic system. In this case the kinetic energy is fixed
$K(\zG)=K_{0}=(dN-d-1)k_{B}T=(dN-d-1)/
\beta$. Equation (\ref{H0dot})
can therefore be used to show,

\be 
\za(\zG)=-\frac{({\dot H}_0(\zG)+\textbf{J}(\zG)V\cdot\textbf{F}_e)
\beta}{dN-d-1} \label{alphaJK0} 
\ee
while the phase space expansion
rate $\zL$ is still expressed by (\ref{LambdaE0}) to order N, with
a different $O_{N}(1)$ term. The equilibrium phase space distribution
function takes the form $f(\Gamma)\sim e^{-
\beta H_{0}}\delta(K(\Gamma)-K_{0})$,
on a surface $K(\zG)=$const, hence $f(\Gamma)/f(iS^{\zt}\Gamma)=e^{
\beta\int_{0}^{\zt}{\dot{H}}_{0}(\zG(s))ds}$,
and Eq.(\ref{omegat}) leads to: 
\be 
\overline{\Omega}_{0,\zt}(\zG)=
\beta \int_0^\zt{\dot H}_0(\zG(s))ds -\overline{\Lambda}_{0,\zt}(\zG)=
-\overline{(\textbf{J}\cdot\textbf{F}_e)}_{0,\zt} V 
\beta +O_N(1) 
\ee
for all $\zt$. Combining this with Eqs.(\ref{alphaJK0}) and (\ref{LambdaE0}),
one obtains 
\be 
\zS(\zG)=-\textbf{J}(\zG)\cdot \textbf{F}_e V
\beta +O_N(1) ~. \label{OmegaJK0} 
\ee

\item Nos\'{e}-Hoover thermostat. In this case, equations (\ref{geneqnsmotion})
are supplemented with the equation, 
\be 
{\dot \za}=\frac{1}{Q}(2K(\zG)-dNk_BT)
\ee
where Q is a constant related to the relaxation time of the thermostat,
and $T$ is the imposed average temperature. In this case, 
\be 
\za(\zG)=-({\dot
H}_0(\zG)+ \textbf{J}(\zG)V\cdot\textbf{F}_e)/(2K(\zG)) \label{alphaJNH}
\ee
and 

\be
\zL(\zG) = \nabla \cdot \dot{\Gamma}+\frac{\partial{\dot \za}}{\partial \alpha}
\stackrel{\rm AI\zG}{=} -dN\za ~.
\label{LambdaNH}
\ee

The equilibrium phase space distribution function is the Nos\'{e}-Hoover
extended canonical distribution $f(\Gamma)\sim e^{-\beta(H_{0}+\frac{1}{2}Q\za^{2})}$,
hence $f(\Gamma)/f(iS^{\zt}\Gamma)=e^{\beta\int_{0}^{\zt}({\dot{H}}_{0}\zG(s)+Q\za\dot{\za})ds}$.
So equation (\ref{omegat}) can be used to show that: 
\be 
\overline{\zS}_{0,\zt}^{NH}(\zG)=
\beta \int_0^\zt({\dot H}_0(s)+Q\za\dot{\za})ds -\overline{\Lambda}_{0,\zt}(\zG)=
-\overline{(\textbf{J}\cdot\textbf{F}_e)}_{0,\zt} V 
\beta 
\ee
for
all $\zt$, and then 
\be 
\zS(\zG)=-\textbf{J}(\zG)\cdot\textbf{F}_e
V \beta \label{OmegaJNH} 
\ee

\end{itemize}
Therefore, for field driven nonequilibrium systems whose equations
of motion satisfy $AI\zG$, the dissipation function and the dissipative
flux are simply related. In fact, for thermostatted systems with a
constant field, $\zS(\zG)=kJ(\zG)$ where $k$ is a constant.

It is straightforward to show that the same expressions (\ref{OmegaJE0}),
(\ref{OmegaJK0}),(\ref{OmegaJNH}) can be obtained even if only a
fraction of the system's particles is subjected to a thermostat (e.g.
the wall particles), see \cite{SE2000},\cite{HeatFlow} for example.
Furthermore, if the walls are large, the results are not sensitive
to the details of the thermostatting mechanism \cite{StephenPRE}.

\section*{APPENDIX B}

The derivation of the steady state relations from transient
ones needs a careful analysis of the conditional average
\be
\left\langle e^{-\zW_{0,t_0}} \cdot e^{ - \zW_{t_0+\zt,2t_0+\zt}}
\right\rangle_{\Wt \in A^+_\zd} 
\label{B1}
\ee
which appears in Eqs.(\ref{exactFT_t_0},\ref{exactFTt_01}), and
of its limit 
\be
M(A,\zd,\zt) = \lim_{t_0 \to \infty}
\left\langle e^{-\zW_{0,t_0}} \cdot e^{ - \zW_{t_0+\zt,2t_0+\zt}}
\right\rangle_{\Wt \in A^+_\zd}
\label{BM}
\ee
which is needed for the transient states $\mu_{t_0}$ to eventually reach the 
steady state $\mu_\infty$. In particular, $M(A,\zd,\zt)$ must exist and be 
positive for $\zt$ larger than a certain $\zt_{A,\zd}$, for the pair $(-A,A)$ 
to be $\zd$-possible, 
i.e.\ for fluctuations of size close to $A$ to occur, and for $A$ to be in 
the domain of the fluctuation relation with some tolerance $\zg > \zd$. 

As observed in the text, Eqs.(\ref{exactFT_t_0}) and (\ref{exactFTt_01}) are exact, 
hence cannot contradict any correct result on deterministic, reversible dynamical 
systems. The question is whether one can extract sufficient information from them, 
that they can be of practical use for the description of the statistics of steady state trajectories. This requires that either the dynamics of
interest be explicitly given, or that the possible situations be explored.
In section 4, we have considered the possibility that the $\zW$-autocorrelation
decays in time, as the most common for the deterministic reversible particle models 
of nonequilibrium fluids. In fact, less is needed for the steady state $\zW$-FR
to hold for $A$; it suffices that $M(A,\zd,\zt)$ does not grow too fast with 
$\zt$: subexponential growths, and exponential growths with rate not 
larger than $\zd$ (as for $A=0$) are all acceptable.

In the linear regime, the decay of correlations in the steady state can be well approximated by their decay at equilibrium, and for this reason, in the linear regime, the transport coefficients are given by the integral of the equilibrium time correlation function, multiplied by the field  \cite{DenisGarybook}.  Lack of decay of correlations of $\zW$ corresponds to the non-existence of the transport coefficient 
associated with $\zW$.  Further away from equilibrium, nonlinear response theory applies.  Now the time correlation function is still determined with respect to the initial measure $\mu$, but the dynamics used to compute the nonlinear transport coefficients is the nonequilibrium one 
(cf.\ the Transient Time Correlation Function \cite{DenisGarybook}).  This is analogous to what is done in calculations of $M(A,\zd,\zt)$, and if the nonlinear transport coefficients exist in the nonlinear regime, then there should be a decay of the $\zW$-autocorrelation, with respect to
the equilibrium measure.
Note also that the growth of $\zt$ increases the separation of  the end of the integral 
in the first exponential of Eq.(\ref{B1}) from the beginning of the integral in 
the second exponential, hence that the growth of $\zt$ should contribute to the 
decorrelation.  

Therefore, it looks quite plausible that, in most cases, one of the above considerations 
on the decay of correlations be satisfied. There is, however, one competing effect,
which may become important in particular circumstances: the set of trajectories 
over which the conditional average (\ref{B1}) is computed changes with $\zt$, 
and this may balance the decorrelation phenomena in some cases.

To illustrate these facts, consider a simple model that has been discussed in connection to the fluctuation theorems in the past \cite{CG99}, and apply the procedure described in this manuscript to that example. The model consists of a particle moving in empty space, under the action of a constant 
external force ${\bf F_e}$ and a Gaussian thermostat: 
\bea
&&\dot\q = \p \nonumber \\
&&\dot\p = {\bf F}_e - \frac{{\bf F}_e \cdot \p}{\p \cdot \p} \p  \nonumber
\eea
where the kinetic energy $K={\bf p \cdot p} / 2$ is a constant of motion. The dynamics of this particle 
is rather simple:
for initial conditions with $\p$ pointing in the direction opposite to ${\bf F}_e$,
$\p$ is a constant of motion; for all other initial conditions the direction of $\p$ 
tends to the direction of ${\bf F_e}$, while the magnitude of $\p$, $\sqrt{2K}$, is  
constant. The treatment of Sections 2 and 3 applies to this system, since it 
is reversible and an initial distribution that is ergodically consistent with the final distribution can be selected.  
However there are various reasons why the treatment in Section 4 cannot be applied to produce a steady state $\zS$-FR in this case. For instance, the distribution generated by the field free equations of motion is not ergodically consistent with the steady state dynamics (see Appendix A).  In addition, the attractor is a fixed point in ${\bf p}$-space, and the time-dependence of $\bf q$ is 
irrelevant; so, for any $\Omega$ that is a function of $\bf p$ only, there are no steady state fluctuations - that is for small $\zd$, no pair $(-A,A)$ 
is $\zd$-possible, hence that the domain of the steady state \WFR is empty. 
The theory outlined in this paper also anticipates these facts, since correlations 
for any phase function do not decay, and therefore the term $M$ 
diverges in the $t_0 \to \infty$ limit.
In particular, selecting the dissipation function that would correspond to a uniform initial distribution, we obtain $\zW=-\zL=(d -1){\bf F}_e \cdot \p / 2K$ where $d$ is the number of dimensions in configuration space. The range of values of 
this $\zW$ is $[-(d-1) \sqrt{2K} | {\bf F_e} | , (d-1) \sqrt{2K} | {\bf F_e} |]$ and any smooth 
probability distribution in it (e.g. the uniform one) evolves in time in 
such a way that 
\be
\mu_{t_0}(\zW_{0,\zt}|_{A_\zd^+}) \stackrel{t_0 \to \infty}{\longrightarrow} 
\left\{ \begin{array}{ll} 
0 & \mbox{if~~ } (d-1) \sqrt{2K} | {\bf F_e} | \notin (A-\zd,A+\zd) \\
 & \\
1 & \mbox{if~~ } (d-1) \sqrt{2K} | {\bf F_e} | \in (A-\zd,A+\zd) \end{array}
 \right.
\ee
As all trajectories, except those with initial condition 
$\p(0)=-(d-1)\sqrt{2K}{\bf F}_e/|{\bf F}_e|$, tend to have 
$\zW=(d-1) \sqrt{2K}{\bf F}_e/| {\bf F_e} |$, this means that the integrals 
$\zW_{0,t_0}$ and $\zW_{t_0+\zt,2t_0+\zt}$ diverge linearly with $t_0$. 

Therefore, Eq.(\ref{aveBDD}) does not apply in this case.  This will be the case for any steady state that has no fluctuations.
Nevertheless, the transient and asymptotic relations remain valid for the 
evolving ensembles. The transient \WFR expresses the 
rate at which the initial probability of observing $-A$ vanishes, compared to 
the initial probability of $A$, and the $\zW_\infty$-FR expresses the corresponding 
asymptotic rate; but the domain of the steady state \WFR is empty.
A similar situation, with no fluctuations
of opposite sign in the steady state, may be produced in more realistic models by applying extremely high fields, making the motion ordered. In these cases, the domain of the steady state \WFR would also be empty.   
However, intermediate situations are possible, like those of Ref.\cite{stephennew}, in which the attractors have small dimension, but
the initial correlations decay, and the steady state \WFR holds. 

References \cite{axiomC,BGG} consider systems with steady states confined within manifolds $\mathcal{M}_r$ of dimension smaller than $\mathcal{M}$, whose volumes fluctuate in time, so that the corresponding phase space contraction rate, $\Lambda_r$, obeys the standard fluctuation relation. In these systems, correlations of all observables decay with respect to the invariant measure $\mu_\infty$, supported on $\mathcal{M}_r$. Furthermore, if the phase space contraction rate $\zL$, in the space of the field-free dynamics $\mathcal{M}$, is proportional to $\zL_r$, as assumed in \cite{BGG}, a modified FR holds for it, which can be expressed as
\be
c A - \zg \le \frac{1}{\zt}
\ln \frac{\mu_{\infty}(\overline{\zL}_{0,\zt}|_{A^+_\zd})}
{\mu_{\infty}(\overline{\zL}_{0,\zt}|_{A^-_\zd})} \le c A + \zg
\label{BaxC}
\ee
with $c<1$.
The relevance of such a possibility for particle systems is not obvious,
but no difficulty emerges with the theory developed here, since the decay of correlations
with respect to $\mu_\infty$ does not imply the decay of correlations with respect to 
a non-singular measure supported on $\mathcal{M}$.
Many different scenarios are possible, and should be analyzed case by case, but this
goes beyond the purpose of the present paper. Here it suffices to observe that no 
scenario that can be realized can contradict the framework of the present paper; 
to the contrary, it could be explained in it, through the analysis of the conditional 
average (\ref{B1}).

\end{document}